\documentclass[11pt]{article}
\pdfoutput=1
\usepackage{jheppub}
\usepackage{amssymb,amsfonts}
\usepackage{mathrsfs}
\usepackage{slashed}
\usepackage{tikz}
\usepackage{tqft}
\usepackage{graphics}
\let\savenumberline\numberline
\def\numberline#1{\savenumberline{#1.}}
\makeatletter
\renewcommand{\@seccntformat}[1]{\csname the#1\endcsname.\,\,}
\makeatother

\newcommand{\CM}{{\cal M}}

\renewcommand{\tilde}[1]{\widetilde{#1}}
\renewcommand{\hat}[1]{\widehat{#1}}
\newcommand{\be}{\begin{equation}}
\newcommand{\ee}{\end{equation}}
\newcommand{\bea}{\begin{eqnarray}}
\newcommand{\eea}{\end{eqnarray}}

\newcommand{\tr}{\textrm{tr}}

\newcommand\secref[1]{{\S\ref{#1}}}
\newcommand\appref[1]{{Appendix~\ref{#1}}}

\def\nn{\nonumber}
\makeatletter
\def\@fpheader{\relax}
\makeatother
\usepackage{graphicx}
\usepackage{latexsym}

\title{Tropical BF Theory and Tropical Limits of TQFTs}
\author{Emil Albrychiewicz and Andr\'{e}s Franco Valiente}
\affiliation{\medskip
Leinweber Institute for Theoretical Physics and Department of Physics\\
University of California, Berkeley, CA, 94720-7300, USA\medskip\\
Theoretical Physics Group, Lawrence Berkeley National Laboratory\\
Berkeley, CA 94720-8162, USA}
\emailAdd{ealbrych@berkeley.edu}
\emailAdd{andresfranco@berkeley.edu}
\abstract{We study anisotropic scaling limits of topological field theories using tropical geometry. The resulting topological field theories are characterized by foliated geometries and are invariant under foliation-preserving gauge transformations. We demonstrate the tropicalization for the 2D BF theory and generalize the prescription to topological Yang-Mills and Chern-Simons theories. We call the tropical limit of the BF theory, the \textit{TBF} theory, which is an anisotropic generalization of the BF theory with an additional adjoint-valued field $T$ that enforces a projectability condition onto the leaves of the foliation. The TBF theory localizes onto the moduli space of tropicalized flat connections $\mathcal{M}(\Sigma_g,G)$ on a foliated Riemann surface $\Sigma_g$ of genus $g$. The tropical connections exhibit anisotropic behavior; their holonomy is sensitive only to the leaves of the foliation. We analyze this moduli space two distinct ways,  Firstly, they are classified by leaf-wise holonomy whose dimension can be explicitly calculated for the case of tropical projective space $\mathbb{TP}^1$ by the moduli space isomorphism $\mathcal{M}\left(\mathbb{TP} ^1, G\right) \cong \operatorname{Hom}(\mathbb{Z}, G) / G$. The second way is through Kodaira-Spencer theory which gives a twisted cohomology argument to argue that $\operatorname{dim} \mathcal{M}\left(\mathbb{T} P^1, G\right)=\operatorname{rank}(\mathfrak{g})$ and we demonstrate their equivalence for the case of SU$(N)$.  We show that we can glue together several $\mathbb{TP}^1$ to obtain $\operatorname{dim} \mathcal{M}\left(\Sigma_g, G\right)=(g-1)\operatorname{rank}(\mathfrak{g})$ for $g \geq 2$ which is precisely $\frac{1}{2}$ of the usual result through an application of a foliated refinement of the Atiyah-Segal axioms. This suggests that a naive tropicalization of a topological field theory will in general not preserve the correlators. We leave several open questions such as potential connections to JT gravity and anisotropic conformal field theory.
}

\begin{document}
\maketitle

\section{Introduction}
Topological field theories, as reviewed in \cite{Birmingham:1991ty}, have long provided deep insights into the nonperturbative structure of quantum field theories, serving as both a robust mathematical framework and a practical tool in condensed matter physics. These theories, by virtue of their metric independence, capture global topological features that are invariant under continuous deformations, making them ideal for studying phenomena such as topological phases of matter and quantum invariants. However, many physical systems, ranging from strongly correlated materials to nematic liquid crystals exhibit inherently non-relativistic anisotropic properties that are not fully captured by traditional, isotropic relativistic field theories.

Anisotropy in field theory is of keen interest because it unlocks a deeper understanding of systems where direction-dependent interactions and scaling may dominate. Lifshitz field theories \cite{lifshitz1941, hornreich1975,chaikin1995} are among the first to capture such anisotropic behavior by introducing different scaling behaviors between time and space and have consequently provided invaluable insights into quantum criticality and non-relativistic systems in condensed matter physics for several decades. Since then, there have been Lifshitz-like generalizations that appeared. In particular, fractonic \cite{Pretko_2018,Pretko_2020,Nandkishore_2019} and foliated field theories \cite{Slagle_2019}, as well as exotic field theories \cite{Gorantla_2020,Gorantla_2022,Seiberg_2021}, which have emerged as interesting field theories that not only describe exotic phases of matter, such as fractons, which have anisotropic behavior and restricted mobility but also give additional insight into subsystem symmetries and higher-form symmetries \cite{Gaiotto_2015}. The relevance of anisotropy reaches into gravitational theories, notably in Hořava-Lifshitz gravity \cite{horava2008}, where similar anisotropic scaling principles offer promising avenues to reconcile quantum mechanics with gravity. 

Recently, it was shown that it is possible to construct explicit examples of anisotropic generalizations of cohomological field theories by applying tropical geometry to topological sigma models, resulting in what is known as a \textit{tropological sigma model}. The question of how tropical geometry can be implemented directly into path integrals was first discussed in \cite{Albrychiewicz:2023ngk} and has had few recent developments in \cite{Albrychiewicz:2024tqe, Albrychiewicz:2025hzt, Albrychiewicz:2025rkg}. It was discovered that tropical geometries can be described by path integrals by using what is known as the Maslov dequantization procedure \cite{msintro, litvinov, viro, virohyper} which is a formal deformation that takes complex varieties and degenerates them into an underlying real algebraic geometry known as a tropical variety. For the case of Riemann surfaces, one can intuitively see that this degeneration reduces a complex geometry into a tropical graph. Unlike the currently accepted perspective on tropical geometry, where one works with a real algebraic geometry, the Maslov dequantization limit of path integrals allows one to represent the underlying tropical geometry as a foliated complex geometry with an additional equivariance condition that allows one to recover the tropical curve. This allows us to leverage all the standard analytic methods of differential geometry in addition to the algebraic combinatorial tools that tropical geometry provides.

Using this perspective, it was shown that the tropical version of topological sigma models recovers the computation of the Gromov-Witten invariants \cite{gromov, ewtsm}, giving a path integral formulation of Mikhalkin's \cite{mikhalkin} results that psuedoholomorphic curves can be obtained by combinatorial tropical methods. An important observation was that the class of worldsheets that arise from the tropical limit are anisotropic in nature, in contrast to the standard relativistic worldsheet theories that describe topological sigma models. One might wonder if the applicability of topological methods can be extended to a wider class of physical systems upon allowing anisotropic behaviors along the same lines of the tropical topological sigma model.

Consequently, in this work, we introduce an tropical version of topological field theory that allows one to still use the framework of isotropic topological field theory while incorporating direction-dependent features through the use of tropical geometry.  In order to answer the above question, it would be insightful to find a simple topological field theory  that can act as a toy model that allows an easy extension of tropical path integrals into different contexts. An elegant choice is the Schwarz-type topological field theory known as BF theory \cite{Schwarz:1978cn, Blau:1989bq, Witten:1991we,Horowitz:1989ng}  In recent decades, BF theory has been extensively studied as a central topological quantum field theory due to its one-loop exactness and localization properties onto the moduli space of flat connections, enabling explicit computations of topological invariants like the Ray–Singer torsion. In three dimensions, BF theory can be explicitly reduced to Chern–Simons theory \cite{Witten:1988hf} upon suitable constraints. Additionally, BF theory naturally generalizes to higher gauge structures, giving rise to extended observables such as Wilson surfaces and accommodating categorical symmetry structures. It also serves as a cornerstone in various formulations of gravitational physics, notably in the BF- Plebanski formulation of general relativity \cite{Gielen_2010}, spin foam models \cite{baez1999introductionspinfoammodels} and the first order formulation of JT gravity \cite{Turiaci:2024cad}. Apart from theoretical considerations, there are more tangible condensed matter systems where the BF theory is relevant such as topological insulators \cite{Cho:2010rk} and the quantum spin Hall effect \cite{Putrov:2016qdo}.  Consequently, in this paper, we will carry out the tropicalization of the two-dimensional BF theory and use the construction to generalize it to other topological field theories.

In \secref{sec:Maslov}, we will review the basics of how tropical geometry arises in quantum field theories by recalling its origins in string theory. We will then review the basics of the BF theory and its connection to the Ray-Singer analytic torsion and the moduli space of flat connections. In \secref{sec:Moduli}, we discuss how tropical geometry can be encoded into conventional path integrals and, for the sake of simplicity, apply this to the two-dimensional BF theory on complex projective space $\mathbb{CP}^1$. We are able to construct the Lagrangian action for the tropical BF theory, which we call the\textit{ TBF theory} by using a Hubbard-Stratonovich transformation and find that its equations of motion impose a projectability condition and a transverse flatness condition, generalizing the standard flatness conditions of the BF theory. After the tropical limit, the complex projective space $\mathbb{CP}^1$ is replaced by the tropical projective space $\mathbb{TP}^1$ but now interpreted as a foliated complex manifold. Further, we investigate the moduli space of tropicalized flat connections and find that it is naturally preserved under foliation-preserving gauge transformations. Using this insight, we compute the dimension of the moduli space of tropicalized flat connections by employing Kodaira-Spencer theory and showing through the twisted cohomology of the deformation complex, that the dimension of the moduli space of tropicalized flat connections for 2-dimensional TBF theory on $\mathbb{TP}^1$ is given by the rank of its gauge group, away from singular irreducible tropical connections. We then demonstrate that we are able to compute the dimension of this moduli space by looking at holonomies that have a natural decomposition along the leaves of the foliation and transverse to the leaves of the foliation. We find that the tropical holonomy is entirely encoded along the leaves of the foliation which allows us to verify the answer that we obtained for the dimension of the moduli space via twisted cohomology. Despite this being calculated for the case of $\mathbb{TP}^1$, we argue that the result also holds locally on a sleeve which can be glued together into foliated Riemann surfaces of arbitrary genus $g\geq 2$ through a refinement of the Atiyah-Segal axioms to obtain that the dimension of the moduli space of tropicalized flat connections on a foliated Riemann surface of genus $g$ is $\operatorname{dim} \mathcal{M}\left(\Sigma_g, G\right)=(g-1) \operatorname{rank}(\mathfrak{g})$. In \secref{sec:ExamplesTBF}, we provide explicit checks of this correspondence between the twisted cohomology arguments and holonomy arguments on a sleeve for the case when the gauge group is SU$(N)$.  

In \secref{sec:Glossary}, we generalize our prescription to several other topological field theories such as 2-dimensional topological Yang-Mills, higher dimensional BF theories, Chern-Simons theories and their associated boundary theories and Wess-Zumino-Witten models \cite{Elitzur:1989nr}. In \secref{sec:Conclusions}, we discuss an extensive list of leftover open questions that are available for quick followups, in particular, how the TBF theory is suggestive of a tropical formulation of JT gravity as well as anisotropic conformal field theory.

\section{The Maslov Dequantization of BF Theory}
\label{sec:Maslov}

In this section, we will review how tropical geometry has first appeared in physics through string theory and how this is connected to the Maslov dequantization procedure which allows one to recover tropical geometries from complex geometries. If one wishes to skip the motivational preamble, the main content of the Maslov dequantization procedure is encoded in \eqref{eqn:Maslov}. We will then review the bare essentials of BF theory and then proceed to demonstrate how the Maslov dequantization procedure can be straightforwardly implemented in the case of the BF theory giving a novel tropical limit of topological field theory with subsystem symmetries which we call the \textit{TBF} theory.

\subsection{Elements of Tropical Geometry in Path Integrals}

In the last few decades, tropical geometry has emerged as a vibrant and unifying framework that translates nonlinear problems of classical algebraic geometry into the combinatorial language of piecewise-linear structures. At its core, tropical geometry replaces the usual operations of addition and multiplication with maximization and addition, respectively, often simplifying calculations. Beyond its role in mathematics, tropical geometry has demonstrated its versatility in other domains. In machine learning, and more generally, computer science, tropical methods have been instrumental in designing efficient algorithms and in understanding the structure of solution spaces in dynamical programming. These cross-disciplinary successes underscore the intrinsic value of tropical techniques: they provide a universal language that bridges the gap between discrete and continuous mathematics, offering fresh perspectives and new computational strategies.

The origins of tropical geometry in physics can be traced back to BPS objects that exist within superstring and M-theory known as string networks. These string networks can be constructed as bound states of half-BPS strings labeled by two co-prime integers $(p,q)$, referred to as $(p,q)$-strings where $p$ and $q$ are the charges under the two distinct two-form gauge fields that exist in the target space supergravity description of Type IIB superstring theory in $\mathbb{R}^{10}$. At weak coupling, one can interpret the $(1,0)$-string as a fundamental string and the $(0,1)$ string as the corresponding Dirichlet string upon which a fundamental string should be allowed to end. As it turns out \cite{aharony1998webs, sni, snii, sniii, sniv}, the $(p,q)$-strings satisfy a conservation law at the string juncture i.e., the total $p$ and $q$ charges must be conserved at each vertex
\begin{equation}
\sum p_i=0, \quad \sum q_i=0.
\end{equation}
This is precisely the matching condition of tropical graphs. One can show that not only must the charges be conserved but the strings must also come under specified fixed angles within a two-plane for supersymmetry to be preserved. Repeatedly fusing $(p,q)$-strings using the matching condition at each vertex gives a piece-wise linear geometry which is a BPS object known as a string network. Mathematically, the resulting object is a tropical curve in two-dimensional tropical projective space $\mathbb{TP}^2$. The crucial insight is that one is able to lift this Type IIB string network to M-theory by using the duality statement that Type IIB superstring theory on $\mathbb{R}^{10}$ is dual to M-theory compactified on a two-torus $T^2$. Employing this duality shows that all the different $(p,q)$-strings arise from the M2 brane and the string network reduces down to a single condition defining a smooth holomorphic embedding of the M2-brane into M-theory compactified on the two-torus $T^2$. In the limit that one of the radii of this torus vanishes, one recovers the string network but clearly sees that the tropical geometry is produced from the holomorphic embedding through a Maslov dequantization limit. The main lesson being that, from the perspective of M-theory, it is much more insightful to keep the additional $S^1$ that one dimensionally reduces on the torus as an auxiliary foliation instead of dropping it altogether. 

In order to employ the lessons above, we view all tropical geometries as being the analog of the string networks and we would like to holomorphically lift it up to a complex geometry where we retain the information of the extra $S^1$, which we call the tropical circle. In the Maslov dequantization limit/tropical limit, we take the radius of this circle to zero to recover the tropical geometry. In practice, this means if we have local complex coordinates $(z,\bar{z}) $ on a complex manifold $\hat\Sigma$, we use hats to distinguish geometric objects before the limit, we can parametrize them as
\begin{equation}
\label{eqn:Maslov}
z=e^{\frac{r}{\hbar}+i \theta}, \quad \bar{z}=e^{\frac{r}{\hbar}-i \theta},
\end{equation}
and then recover the underlying tropical geometry $\Sigma$ in the limit where $\hbar\rightarrow0$. In doing so, we keep the differential structure of the complex geometry and instead deform the geometric structures defined on the complex structure $\hat{\Sigma}$ . In particular, the complex structure $\hat{J}$ defined on the Riemann surface degenerates into a nilpotent endomorphism on the tangent bundle $J$. In tandem, this then defines a distribution on $\Sigma$ which is automatically integrable in the two-dimensional case and hence defines a foliation on $\Sigma$. As a result, tropical geometries can be recovered in the tropical limit i.e., $\hbar\rightarrow 0$ as a foliated complex geometry.

Given the fact that we now have a foliation, we have a privileged coordinate system known as adapted coordinates, denoted $(r,\theta)$, in which the Jordan structure takes the form
\begin{equation}
J=\left[\begin{array}{ll}
0 & 1 \\
0 & 0
\end{array}\right].
\end{equation}
In the tropical limit, the metric tensor is now a degenerate bilinear form that in adapted coordinates has the matrix representation
\begin{equation}
g  =\left[\begin{array}{ll}
1 & 0 \\
0 & 0
\end{array}\right].
\end{equation}
Interpreting tropical geometries in this way allows us to avoid the problem of having to formulate a path integral using the tropical semiring which formally replaces the addition and multiplication of $\mathbb{R}$ with mini-max operations and additions on $\mathbb{T}$ respectively. Instead, we can formulate tropical path integrals using conventional complex numbers and retain all our usual differential methods, albeit now adapted to manifolds with possibly singular foliations. It can be shown that one can extend this Maslov dequantization limit to odd-dimensional geometries as well.

One finds that the nilpotency of the Jordan structure $J$ is preserved under foliation-preserving diffeomorphisms
\begin{equation}
\begin{aligned}
& \widetilde{r}=\widetilde{r}(r), \\
& \widetilde{\theta}=\widetilde{\theta}_0(r)+\theta \partial_r \widetilde{r}(r).
\end{aligned}
\end{equation}
By looking at the Lie algebra associated to this symmetry group, we obtain
\begin{equation}
\begin{aligned}
\delta r & =f(r), \\
\delta \theta & =F(r)+\theta \partial_r f(r).
\end{aligned}
\end{equation}
Thus, the local symmetries of the Jordan structure are generated by the infinite-dimensional Lie algebra whose elements are parametrized by two real, arbitrary, projectable, and differentiable functions $f(r)$ and $F(r)$ on the foliation. Projectability implies that it is leafwise constant, i.e., a basic function. This is in contrast to Riemann surfaces without foliations where we can use the full set of diffemorphisms and not be restricted to foliation-preserving diffeomorphisms.

We want to emphasize that we preserve the differential structure of the complex manifold $\hat{\Sigma}$ upon taking the tropical limit and consequently only deform geometric structures such as complex structures, metric tensors, connections and so forth. Vector bundles defined over the tropicalized manifold $\Sigma$ now have transition functions whose cocycle conditions are defined using foliation-preserving diffeomorphism but otherwise, no component of any section is arbitrarily taken to vanish. We allow for rescaling/renormalization of sections and other physical constants.

\subsection{Elements of BF Theory }
Now that we have reviewed the foliated complex geometry perspective on tropical geometry provided by \cite{Albrychiewicz:2023ngk}, we will proceed to reviewing the essentials elements of BF theory. 

BF theory is a Schwarz-type topological gauge theory defined on a arbitrary $n$-dimensional manifold $M$. The name arises from its field content where $B$ is a $(n-2)$ adjoint-valued differential form and $F$ is a adjoint valued differential 2-form known as the Yang-Mills curvature associated to the gauge group $G$. The Lagrangian action for BF theory can be written as
\begin{equation}
S_{B F}[B, A]=\int_M \operatorname{tr}(B \wedge F),
\end{equation}
where $F=d A+A \wedge A$ and the $\operatorname{tr}$ is the trace taken in the adjoint representation of the Lie algebra $\mathfrak{g}$ of the gauge group $G$.  One can explicitly see that this theory is topological because its action does not depend on any metric and thus can only describe global, topological aspects of the gauge fields. Its equations of emotion are given by the flatness condition
\begin{equation}
F(A)=0,
\end{equation}
and the covariant constancy condition
\begin{equation}
d_A B \equiv d B+[A, B]=0.
\end{equation}
For arbitrary dimensions, one can see that the BF theory is not only manifestly diffeomorphism invariant but also enjoys gauge symmetries
\begin{equation}
\begin{gathered}
\delta_\lambda A=d_A \lambda \equiv d \lambda+[A, \lambda], \\
\delta_\lambda B=[B, \lambda],
\end{gathered}
\end{equation}
and for BF theories that are defined on manifolds whose dimension is greater than two, a topological shift symmetry given by
\begin{equation}
\begin{gathered}
\delta_\eta A=0, \\
\delta_\eta B=d_A \eta \equiv d \eta+[A, \eta].
\end{gathered}
\end{equation}
This additional topological symmetry is a reducible gauge symmetry. As a result, when quantizing using the BF theory using the BRST formalism, one introduces ghost fields for each independent gauge symmetry. Because of the reducibility, the standard ghost fields are not enough to fully gauge fix the symmetry. One must introduce additional ghost fields—often called ``ghosts for ghosts" or secondary generation ghosts—to account for the fact that some gauge variations vanish identically. 

If one temporarily ignores the intricate ghost structure, one can immediately see that the BF partition function effectively localizes onto the moduli space of flat connections $\mathcal{M}(M,G)$. Formally,
\begin{equation}
Z_{BF}=\int \mathscr{D} A\left(\int \mathscr{D} B e^{i \int_M \operatorname{tr}(B \wedge F)}\right)=\int \mathscr{D} A \delta(F)=\operatorname{Vol}(\mathcal{M}(M,G)).
\end{equation}
Consequently, in order to compute the partition function, one would like to do an analysis of the moduli space of flat connections which entails finding an appropriate integration measure over this space. This would entail doing a more detailed analysis of the ghost structure. Generically, this moduli space can be very singular due to the fact that reducible/singular connections may appear which manifest as ghost zero modes. Likewise, the antighosts that come from a prescribed gauge fixing condition may also have zero modes and this can be linked back to the covariant constancy condition having non-trivial solutions \cite{Blau:2022krm}. For the simple case, that we have isolated, irreducible flat connections $A$, the path integral can be shown to be related to the Ray-Singer analytic torsion \cite{RAY1971145}. The Ray-Singer torsion $\tau(M, A)$ on a smooth $d$-dimensional manifold $M$ is a topological invariant that is defined as a product of functional determinants
\begin{align} 
\label{eqn:RSTDef}
\tau(M, A) = \prod_{p=0}^d (\text{det}' \Delta_A^p)^{-(-1)^p \frac{p}{2}}, 
\end{align} 
where $\det'$ indicates that zero modes of the twisted Laplacians were removed. 

Beyond the partition function, there are natural classes of gauge-invariant observables available that one may insert into the path integral. The first class would be Wilson loop observables
\begin{equation}
W_R(\gamma)=\operatorname{tr}_R\left(\mathcal{P} \exp \oint_\gamma A\right),
\end{equation}
where $\gamma$ is a closed loop in the manifold $M$, $R$ is a representation of the gauge group and $\mathcal{P}$ is the path ordering operator. Due to the fact that BF theory localizes onto flat connections, these loops only depend on the homotopy class of $\gamma$ and consequently measure nontrivial holonomies around non-contractible homology cycles. For the case of $\mathbb{CP}^n$ these holonomies would vanish due to spheres being simply connected for $n \geq 1$. We will find that in the case of tropical projective space $\mathbb{TP}^1 $ , we have an additional $S^1$ that stops us from trivializing the path integral.  Higher dimensional formulations of BF theory, in particular,  in 4-dimensions, allows us to interpret the $B$ field as a higher rank gauge field which can then be integrated over a homology 2-cycle and exponentiated into a Wilson surface observable. One could consider inserting into the path integral, the following observable
\begin{equation}
\mathcal{O}(\Sigma)=\operatorname{tr}\left(\exp \int_{\Sigma} B\right).
\end{equation}
Or more simply, factors of the $B$ field as
\begin{equation}
\tilde{\mathcal{O}}(\Sigma)=\int_{\Sigma} \operatorname{tr}(B).
\end{equation}
For the case of a single Wilson loop insertion, one can show that the path integral reduces down to an integration over the moduli space of flat connections with characters associated to the representation $R$ which the Wilson loop is traced over. Schematically,
\begin{equation}
Z_{BF}\left[W_R(\gamma)\right] \propto(\operatorname{dim} R)^{\chi(M)-1},
\end{equation}
here $\chi(M)$ is the Euler characteristic of the manifold $M$.

For several Wilson loop insertions, where each loop $\gamma_i$ is associated to a representation $R_i$ and $i\in\{1,..,k\}$ one finds that the path integral can be decomposed into two inequivalent classes. The first class is the case of non-interacting/unlinked Wilson loops. The path integral effectively factorizes into single Wilson loops
\begin{equation}
Z\left[W_{R_1}\left(\gamma_1\right) \cdots W_{R_n}\left(\gamma_n\right)\right] \propto \prod_{i=1}^n\left(\operatorname{dim} R_i\right)^{\chi(M)-1}.
\end{equation}
The second class is the case of linked Wilson loops where the holonomies are coupled over a common link $U$. In this case, one is able to decompose the corresponding characters $\chi_R$ using a Clebsch-Gordon/fusion decomposition
\begin{equation}
\chi_{R_1}(U) \chi_{R_2}(U)=\sum_{R_3} N_{R_1 R_2}^{R_3} \chi_{R_3}(U),
\end{equation}
where the $N_{R_1 R_2}^{R_3}$ are the fusion coefficients of the decomposition of  $R_1 \otimes R_2$. For the case of two linked Wilson loops, the path integral evaluates to a sum over intermediate representations $R_3$
\begin{equation}
Z\left[W_{R_1} W_{R_2}\right] \propto \sum_{R_3} N_{R_1 R_2}^{R_3}\left(\operatorname{dim} R_3\right)^{\chi(M)-\Delta},
\end{equation}
here $\Delta$ is a shift that depends on the number of Wilson loop insertions. For the case of two-dimensional BF theory, we will not encounter any fusion phenomena since $\mathbb{CP}^1$ is topologically trivial to support linking. In particular, it is simply connected i.e., the fundamental group $\pi_1(\mathbb{CP}^1)$ is trivial.

\subsection{Tropical BF Theory on $\mathbb{TP}^1$}
\label{sec:TropBF}

We now employ the Maslov dequantization limit to the BF theory defined on $\mathbb{CP}^1$. In the standard BF theory, the simply connectedness of $\mathbb{CP}^1$ trivializes the moduli space of flat connections. After tropicalization, $\mathbb{CP}^1$ becomes the tropical projective space $\mathbb{TP}^1$ but now represented as a foliated $\mathbb{CP}^1$. With the points at infinity, we are able to identify $\mathbb{TP}^1$ as a infinite foliated cylinder. However, we will see that the moduli space of tropicalized flat connections is insensitive to the radial direction and the infinite foliated cylinder can be replaced by a local geometry, known as a sleeve \cite{Albrychiewicz:2023ngk}, which can then be glued together in order to make more topologically complex foliated Riemann surface through matching conditions where the foliations meet at a juncture.  We begin with the Lagrangian BF action on $\mathbb{CP}^1$
\begin{equation}
S_{\mathrm{BF}}=\int_{\mathbb{CP}^1} \operatorname{Tr}(B F(A)).
\end{equation}
In this case, the $B$ field is an adjoint-valued 0-form scalar field. We employ local complex coordinates on this space such that the curvature two-form takes the following form
\begin{equation}
F(A)=\left(\partial_z A_{\bar{z}}-\partial_{\bar{z}} A_z+\left[A_z, A_{\bar{z}}\right]\right) d z \wedge d \bar{z}.
\end{equation}
The Maslov dequantization of coordinates is given by \eqref{eqn:Maslov},
\begin{equation}
z=e^{\frac{r}{\hbar}+i \theta}, \quad \bar{z}=e^{\frac{r}{\hbar}-i \theta}. 
\end{equation}
The differential forms dz and d$\bar{z}$ transform as
\begin{equation}
d z=z\left(\frac{d r}{\hbar}+i d \theta\right), \quad d \bar{z}=\bar{z}\left(\frac{d r}{\hbar}-i d \theta\right) .
\end{equation}
The corresponding volume form is then
\begin{equation}
d z \wedge d \bar{z}=\frac{-2 i}{\hbar} e^{2 r / \hbar} d r \wedge d \theta.
\end{equation}
The holomorphic and antiholomorphic derivatives and components of one-forms transform according to the tensor transformation law as
\begin{equation}
\begin{aligned}
\partial_z & =e^{-r / \hbar-i \theta}\left(\frac{\hbar}{2} \partial_r-\frac{i}{2} \partial_\theta\right), \\
\partial_{\bar{z}} & =e^{-r / \hbar+i \theta}\left(\frac{\hbar}{2} \partial_r+\frac{i}{2} \partial_\theta\right) , \\
A_z & =e^{-r / \hbar-i \theta}\left(\frac{\hbar}{2} A_r-\frac{i}{2} A_\theta\right) ,\\
A_{\bar{z}} & =e^{-r / \hbar+i \theta}\left(\frac{\hbar}{2} A_r+\frac{i}{2} A_\theta\right).
\end{aligned}
\end{equation}

In order to properly incorporate the tropical limit, we have to impose an additional anisotropic scaling behavior on the radial and angular components of the connection as was done in the construction of tropical topological sigma models. In order to have a convergent limit for the action, we must scale the adjoint zero-form field $B$ in the opposite direction. Altogether, the prescription is
\begin{equation}
\begin{aligned}
A_r & \rightarrow A_r, \\
A_\theta & \rightarrow \hbar A_\theta, \\
B & \rightarrow \frac{1}{\hbar} B.
\end{aligned}
\end{equation}
Asymptotically to $\hbar\rightarrow 0$, our action has the form
\begin{equation}
S_{B F} \sim\int_{\mathbb{CP}^1} \operatorname{Tr}\left(\frac{1}{\hbar} B\left[\hbar\left(\partial_r A_\theta+\left[A_r, A_\theta\right]\right)-\partial_\theta A_r\right]\right)dr\wedge d\theta .
\end{equation}
By performing a Hubbard–Stratonovich transformation such as discussed in the appendix of \cite{Albrychiewicz:2024tqe}, one can directly take the tropical limit $\hbar\rightarrow 0$ by inserting an additional adjoint-valued scalar field $T$ and find that our action has the form
 \begin{equation}
S_{T B F}=\int_{\mathbb{TP}^1} d r d \theta \hspace{0.1cm}\operatorname{Tr}[T\left(-\partial_\theta A_r\right)+B\left(\partial_r A_\theta+\left[A_r, A_\theta\right]\right)
].
\end{equation}

Honoring the naming convention of BF theory, we call the tropical limit of BF theory, the \textit{TBF} theory. Notice that in the tropical limit, the action is now written as an integral over tropical projective space $\mathbb{TP}^1$ now represented as a foliated complex manifold. The foliation arises because $\mathbb{CP}^1$ comes equipped with a canonical complex structure that gets degenerated into a Jordan structure which induces a foliation that decomposes our manifold into equivalence classes given by the leaves of the induced foliation and the space transverse to the leaves of the foliation. The leaves of the foliation are topologically isomorphic to $S^1$. Although, the action does not explicitly depend on this foliation, in the sense that the Jordan structure does not appear, we will see that it has consequences on the transition functions that define any bundle built over this space and consequently, all corresponding sections by restricting the usual diffeomorphisms to foliation-preserving diffeomorphisms. 

The equations of motion for TBF theory is a refinement of the flatness condition of BF theory. We find that $A$ now satisfies a transverse flatness condition
\begin{equation}
f_{r\theta}=\partial_r A_\theta+\left[A_r, A_\theta\right]=0,
\end{equation}
and a projectability condition on the radial component imposing leaf-wise constancy
\begin{equation}
\partial_\theta A_r=0 .
\end{equation}
If $A=A_rdr+A_\theta d\theta$  satisfies both the transverse flatness and projectability condition, we say that $A$ is a tropical flat connection. The transverse flatness condition is no longer described by an antisymmetric two-form. The transverse flatness condition effectively states that $A_\theta$ is covariantly constant along the $r$-direction with respect to the connection $A_r$. In other words, if you fix a ``basepoint" in the $r$-coordinate, then the value of $A_\theta$ at any other point is obtained by parallel transport along $r$ with connection $A_r$. One may think of this as the statement that the ``holonomy" along the $r$-direction preserves $A_\theta$. We will see soon that a manifestation of this statement will be that the holonomies that characterize TBF theory are entirely encoded along the leaves of the foliation in the $\theta$-direction. 

\begin{figure}
    \centering
    \includegraphics[width=0.3\linewidth]{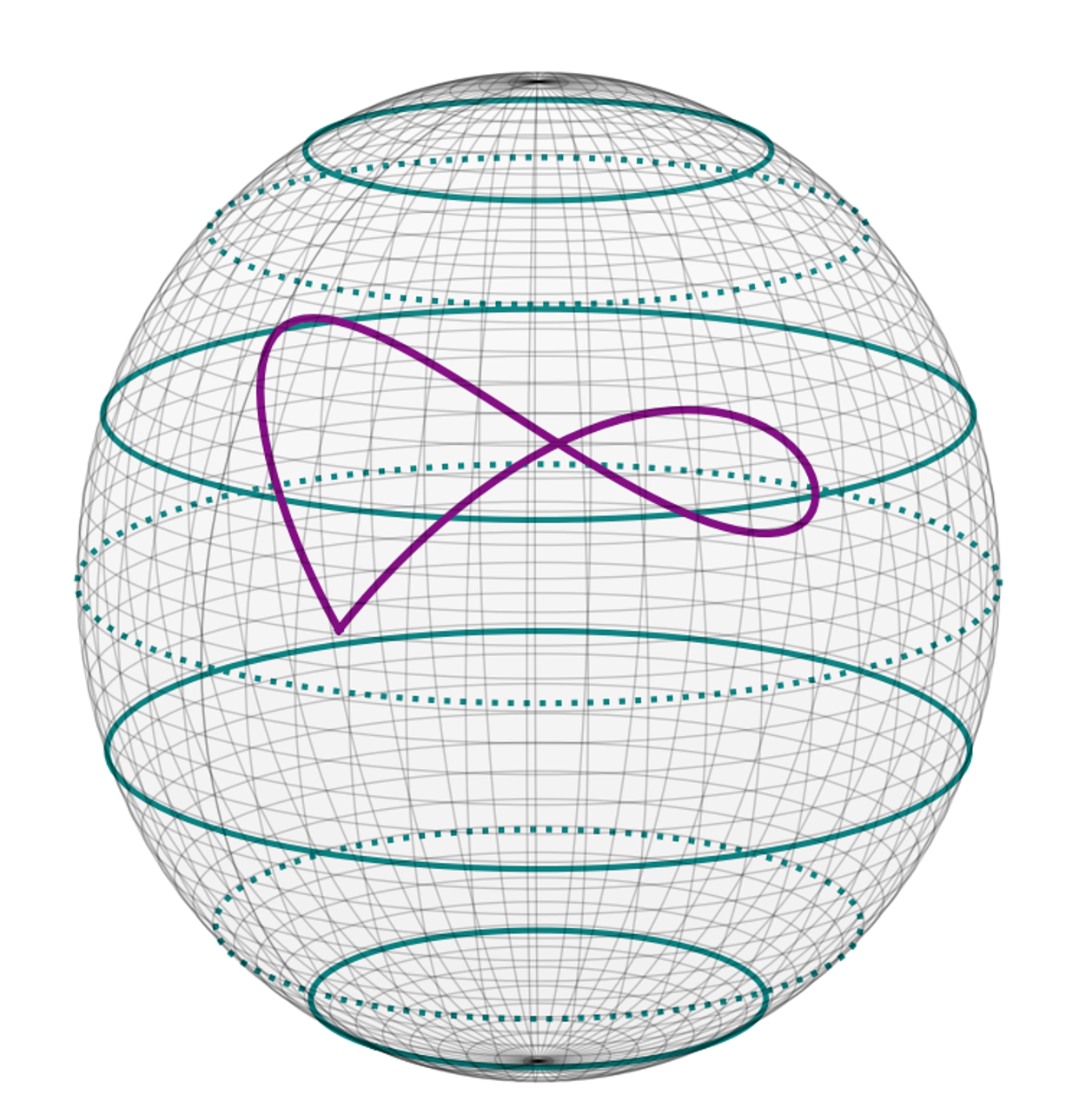}
    \caption{An arbitrary loop $\gamma$ on the foliated $\mathbb{CP}^1 $ that represents the tropical projective space $\mathbb{T}P^1$.}
    \label{fig:enter-label}
\end{figure}

Just like BF theory, we have two additional equations of motion for the $T$ and $B$ fields that enforce covariant constancy. By varying the action with respect to the tropical connection components $A_r, A_\theta$, one obtains
\begin{equation}
\begin{aligned}
& \partial_\theta T+\left[A_\theta, B\right]=0, \\
& \partial_r B+\left[A_r,B \right]=0.
\end{aligned}
\end{equation}
Formally, after integrating out $T$ and $B$, the path integral should reduce down to an integration over the moduli space $\mathcal{M}(\mathbb{TP}^1,G)$ of tropicalized flat connections over tropical complex projective space $\mathbb{TP}^1$ associated to the gauge group $G$. We can write this down as
\begin{equation}
Z_{TBF}=\int_{\mathcal{M}_{\text {}}(\mathbb{TP}^1, G)} \frac{\mathscr{D} A_r\mathscr{D} A_\theta}{\operatorname{Vol}(G)} \Delta_{\mathrm{ghost}} e^{-S[A]},
\end{equation}
where we would have to factor an appropriate volume of the tropicalized gauge group $G$ in order to avoid overcounting of the physical degrees of freedom once we have gauge fixed appropriately, by using either standard cohomological BRST gauge fixing methods or the Faddev-Popov procedure if the gauge symmetries are not too complicated. In order to implement this, we need to characterize the gauge symmetries of TBF theory and its associated localization locus. We will discuss this in \secref{sec:Moduli}.

Before we continue, we want to emphasize that we are only dealing with the TBF theory on an infinite cylinder but will be able to show its equivalence to the TBF theory on a sleeve i.e., a finite foliated cylinder. The question of how to glue these local patches in order to get more topologically complicated foliated Riemann surfaces requires a little bit of engineering guesswork in the following sense: it is not obvious which sort of foliation junctures should be admissible. Naturally, the sort of admissible Jordan structures that one should be allowed to pick should match physical constraints such as finite energy conditions ensuring that the Euclidean action is bounded yet not infinitely suppressed so that relevant configurations still contribute. Such a consistent classification of admissible foliations was developed in the context of tropical topological sigma models is described in \cite{Albrychiewicz:2023ngk} and would be applicable for the case of tropical limit of topological field theories as well. The Jordan structure effectively enters the theory as a non-dynamical background gauge field. Consequently, we expect the path integral of these anisotropic topological field theories to be sensitive to Jordan structure/foliation; however, we do not have to explicitly integrate over all such foliations and address the possible singularities that could arise from exotic junctures. In principle, once an admissible class of foliations is selected, the quantization procedure of the TBF theory could be carried out for a fixed foliation analogously to the BF theory while taking into consideration the gluing/matching conditions at the juncture/vertices. 

In order to get an explicit demonstration of how the matching condition can be implemented at the classical level, we make the observation that imposing divergenceless $\partial_rA_r+\partial_\theta A_\theta=0$ as a preliminary covariant gauge condition recovers a natural notion for what a tropical 1-form should look like. The covariant gauge along with the projectability conditions forces the tropical connection $A$ to have the same structure that one could expect one-forms $\omega$ to exhibit on the edge of a tropical curve. Namely, one-forms $\omega$ should be characterized by a single constant along the $r$-direction of the graph $\omega= \omega^0_rd r$, where $\omega^0_r$ is a constant. Imposing these conditions yields
\begin{equation}
\begin{aligned}
& A_r(r, \theta)=A_r(r), \\
& A_\theta(r, \theta)=A_\theta(r)-\theta \frac{d A_r}{d r}.
\end{aligned}
\end{equation}
Including the periodicity condition of $A_\theta(r, \theta+2 \pi)=A_\theta(r)$, reduces this down to
\begin{equation}
A=A_r^0 d r+A_\theta(r) d \theta,
\end{equation}
here $A^0_r$ is a constant. If one temporarily ``forgets the phase $\theta$" as is done in conventional tropical geometry, we see that we have recovered the mathematical structure for a tropical 1-form. 

If one is interested in gluing several sleeves together in order to get a more topologically complicated foliated Riemann surfaces, the matching condition is then just the sum of all the individual radial zero modes. For the case of three vertices coming together, we then have the matching condition
\begin{equation}
A_{r_<}^0+ A_{r_>}^0+A_{r_\wedge}^0=0,
\end{equation}
where the orientation of the wedge denotes the direction from which the edge is coming into the vertex.  Inserting the solution $A=A_r^0 d r+A_\theta(r) d \theta$ to the equations of motion gives a system of linear ordinary differential equations
\begin{equation}
\partial_r A_\theta^c+i A_r^{a0} A_\theta^b(r) f_{a b}^c=0,
\end{equation}
and we have used the following convention for the structure constants $\left[t_{a,} t_b\right]=if_{a b}^c t_c$. We can organize this by defining the matrix $(M)_b^c \equiv A_r^{a0} f_{a b}^c$, which then allows us to solve this system of ODEs via a matrix exponential, the explicit solution is
\begin{equation}
A_\theta^c(r)=\left[e^{-ir M}\right]_b^c A_\theta^b(0) .
\end{equation}
This suggests that motion along the radial direction simply mixes the phases of the angular component $A_\theta$. For the $U(1)$ case, the transition functions can take into account non-trivial bundles and enhance the periodicity condition by allowing a winding number $w$, namely $A_\theta(r, \theta+2 \pi)=A_\theta(r)+2 \pi w$, if one imposes this, then one notices that the tropical connection has a piece-wise linear structure reminiscent of tropical curves
\begin{equation}
\begin{aligned}
& A_r(r)=A_r^0+w r, \\
& A_\theta(r, \theta)=A_\theta(r)-w \theta.
\end{aligned}
\end{equation}

\section{The Moduli Space of Tropicalized Flat Connections}
\label{sec:Moduli}
In order to characterize the moduli space of tropicalized flat connections, we want to classify which gauge transformations preserve the transverse flatness condition, the projectability condition and the Jordan structure that defines the foliation on the complex geometry. In particular, we expect to discover that we have foliation-preserving gauge transformations. Once, we have the correct notion of gauge transformations, we would like to know if this moduli space is still finite-dimensional so we can explicitly compute the path integral.

Recall that we identified the differential structure of $\mathbb{TP}^1$ to be isomorphic to the differential structure of the Riemann surface $\mathbb{CP}^1$ except we equip complex projective space with a Jordan structure $J$, which is a nilpotent endomorphism of the tangent bundle i.e., $J^2=0$ which induces a foliation. The moduli space of tropicalized flat connections associated to a gauge group $G$ is defined as
\begin{equation}
\mathcal{M}\left(\mathbb{TP}^1, G\right)=\left\{A \in H^1\left(\mathbb{TP} ^1, \operatorname{ad} G\right) \mid \partial_\theta A_r=0, \partial_r A_\theta+\left[A_r, A_\theta\right]=0\right\} / G.
\end{equation}
As previously discussed, the first condition is a projectability condition that tells us that the radial part of the connection is leaf-wise constant, reflecting compatibility with the underlying foliation. We will show that this restricts our diffeomorphisms to be foliation-preserving diffeomorphisms. 

We begin with the transverse flatness condition
\begin{equation}
f_{r\theta}=\partial_r A_\theta+\left[A_r, A_\theta\right]=0.
\end{equation}
Notice here that the object $f_{r\theta}$ is not a standard curvature two-form because it is not antisymmetric. The fact that we are working over a foliated manifold does not change the basic construction of a gauge transformation i.e., they should still be automorphisms of the principal bundle, in particular, for a gauge transformation given by $g$, we have
\begin{equation}
A^g=g^{-1} A g+g^{-1} d g,
\end{equation}
and as well
\begin{align}
    T&=g^{-1}Tg, \\
    B&=g^{-1}Bg. 
\end{align}
If we look at the radial component
\begin{equation}
A_r^g=g^{-1} A_r g+g^{-1} \partial_r g,
\end{equation}
It is clear that the gauge transformed connection will still satisfy the projectability condition $\partial_\theta A^{g}=0$ only if \begin{equation}
\partial_\theta g=0.
\end{equation}
This is the statement that the gauge transformation must be foliation preserving.  Consequently, our gauge transformations reduce down to
\begin{equation}
A_r^g=g(r)^{-1} A_r g(r)+g(r)^{-1} \partial_r g(r),
\end{equation}
\begin{equation}
A_\theta^g=g(r)^{-1} A_\theta g(r).
\end{equation}
Quite intuitively, our restricted class of gauge transformations also preserves the transverse flatness condition
\begin{equation}
f_{r \theta}^g=g^{-1}(r) f_{r \theta} \hspace{0.1cm}g(r).
\end{equation}
Infinitesimally, the foliation-preserving gauge transformations reduce down to
\begin{equation}
\begin{aligned}
& \delta_\lambda A_r=\partial_r \lambda+\left[A_r, \lambda\right], \\
& \delta_\lambda A_\theta=\left[A_\theta, \lambda\right].
\end{aligned}
\end{equation}
The infinitesimal gauge transformations are generated by projectable adjoint-valued scalar fields $\lambda^a(r)$. By restricting our gauge transformation to depend solely on the radial direction, we can see that our gauge symmetry does not act uniformly over the entire $\mathbb{TP}^1$ but instead on a family of 1-dimensional subsystems given by the leaves of the the foliation. This is characteristic of the subsystem symmetries \cite{Seiberg_2021} that emerged in the description of fracton phases in quantum many-body systems \cite{Ohmori_2023}.  Unlike TBF theory which is still a topological field theory, the quantum field theories that describe fractons are generically non-topological quantum field theories that tend to have exotic global symmetries known as subsystem symmetries that restrict the movement of particles, potentially even completely restricting the motion of a particle altogether. It would be interesting to see if there is a topological analog of fracton, potentially existing as an edge mode, that can be characterized through the use of the anisotropic topological field theories. We will postpone this discussion until \secref{sec:Glossary}.

Now that we have identified the restricted gauge of gauge transformations, we would like to understand what classifies inequivalent flat tropical connections. In order to do so, we implement the same procedure that Kodaira and Spencer proposed when deforming complex analytic structures \cite{Kodaira1960ONDO}. We begin with a foliation-preserving deformation of our tropical connection in order to investigate the tangent bundle of the moduli space of tropical connections $T_A \mathcal{M}\left(\mathbb{TP}^1, G\right)$
\begin{equation}
A \rightarrow A+a, \quad \text { with } a=a_r  (r)d r+a_\theta (r,\theta) d \theta.
\end{equation}
We linearize the transverse flatness condition to obtain the linearized tropical flatness conditions that characterizes the deformation complex
\begin{equation}
\partial_r a_\theta+\left[A_r, a_\theta\right]+\left[a_r, A_\theta\right]=0, \quad \partial_\theta a_r=0.
\end{equation}
We can construct the deformation complex by defining the space of gauge parameters as
\begin{equation}
C^0=\left\{\lambda(r) \in \Omega^0\left(\mathbb{T P}^1, \operatorname{ad} G\right) \mid \partial_\theta \lambda=0\right\},
\end{equation}
and the space of tropical connection deformations as
\begin{equation}
C^1=\left\{a=a_r(r) d r+a_\theta(r, \theta) d \theta \mid \partial_\theta a_r=0\right\} .
\end{equation}
We define the differential $d_0: C^0 \rightarrow C^1$ via the linearized gauge transformation
\begin{equation}
d_0 \lambda=\left(\partial_r \lambda+\left[A_r, \lambda\right]\right) d r+\left[A_\theta, \lambda\right] d \theta.
\end{equation}
Then in order to be in the kernel of the next differential in the complex, we impose the linearized flatness condition
\begin{equation}
d_1 a \equiv \partial_r a_\theta+\left[A_r, a_\theta\right]+\left[a_r, A_\theta\right]=0.
\end{equation}
We will show in \appref{app:Nil} that this defines a cohomology theory by explicitly computing $d_1 \circ d_0=0$. The deformation complex is then
\begin{equation}
0 \rightarrow C^0 \xrightarrow{d_0} C^1 \rightarrow 0,
\end{equation}
consequently, we can see the tangent space to the moduli space of tropicalized flat connections is given by the first twisted cohomology group
\begin{equation}
H^1\left(\mathbb{T P}^1, \operatorname{ad} G\right)=\frac{\left\{a \in C^1 \mid d_1 a=0\right\}}{\operatorname{Im}d_0 }.
\end{equation}

As we will argue below, we will be able to gauge fix the tropical connection in such a way that $A_r=0$ and $A_\theta=H$, where $H$ is lying in the Cartan subalgebra $\mathfrak{h} \subset \mathfrak{g}$. Since the gauge parameter $\lambda(r)$ is independent of $\theta$, it can only remove the component of $a_\theta$ that lies in the image of the adjoint action of $H$. In other words, the only deformations not removed by gauge transformations are those lying in the kernel of $\operatorname{ad}_H$, i.e., the Cartan directions. Altogether, one finds that the nontrivial cohomology is isomorphic to the Cartan subalgebra $H^1 \cong \mathfrak{h}$ and therefore
\begin{equation}
\operatorname{dim} \mathcal{M}\left(\mathbb{TP}^1, G\right)=\operatorname{dim} H^1=\operatorname{rank}(\mathfrak{g}).
\end{equation}
Thus, we can see that by imposing the projectability condition, the infinite dimensional space of deformations on the foliated complex geometry reduces everything down to a problem where we effectively only have one residual global degree of freedom which are the constant modes of $A_\theta$, that after conjugation, live in a Cartan subalgebra. We can expect this formula to be true up to singular irreducible connections that might cause a singularity in the moduli space causing the dimension to jump.

Consequently, the path integral of TBF theory generically reduces down to an finite-dimensional integration over the moduli space of tropicalized flat connections. In the non-tropical case, we can explicitly perform the functional integration of BF theory through the isomorphism of the moduli space of flat connections and its associated holonomy group which are classified by gauge group monodromy on the base space or equivalently by group homomorphisms $\pi_1\left(X\right)  \rightarrow G$ up to conjugation. In order to repeat the same construction, we need to see how parallel transport is modified upon tropicalization.

Consider an arbitrary smooth loop $\gamma(t)=(r(t), \theta(t)),  t \in[0,1]$, which satisfies the periodicity condition $\gamma(0)=\gamma(1)$. We can parallel transport a section using our tropical flat connection by solving the following ODE
\begin{equation}
\frac{d \psi}{d t}+\left(\dot{r}(t) A_r(r(t))+\dot{\theta}(t) A_\theta(r(t), \theta(t))\right) \psi(t)=0 .
\end{equation}
The solution to this is the path-ordered exponential
\begin{equation}
\psi(1)=P \exp \left(-\int_0^1\left(\dot{r}(t) A_r(r(t))+\dot{\theta}(t) A_\theta(r(t), \theta(t))\right) d t\right) \psi(0) .
\end{equation}
We can analyze the path-ordered exponential by decomposing the path along the leaves of the foliation $\theta$ and also on transverse segments of constant $r$. For radial segments, where $\dot{\theta}=0$, we have the radial parallel transport operator 
\begin{equation}
U_r\left(r_1, r_0\right)=P \exp \left(-\int_{r_0}^{r_1} A_r\left(r^{\prime}\right) d r^{\prime}\right).
\end{equation}
However, this can be gauged away (for example by using gauge fixing discussed right above) because we can explicitly find a projectable gauge transformation such that transport along the radial directions does not change the section $\psi$
\begin{equation}
U_r^g\left(r_1, r_0\right)=P \exp \left(-\int_{r_0}^{r_1} A_r^g\left(r^{\prime}\right) d r^{\prime}\right)=P \exp \left(-\int_{r_0}^{r_1} 0 d r^{\prime}\right)=\mathbf{1} .
\end{equation}
As a result, the nontrivial information of the parallel transport is entirely contained along the leaves of the foliation. Due to the fact that the radial parallel transport does not contribute to the holonomy, we are allowed to reduce down the infinite foliated cylinder into a finite length foliated cylinder which can then be glued together to form more complicated foliated Riemann surfaces. 
\begin{figure}
    \centering
    \includegraphics[width=0.4\linewidth]{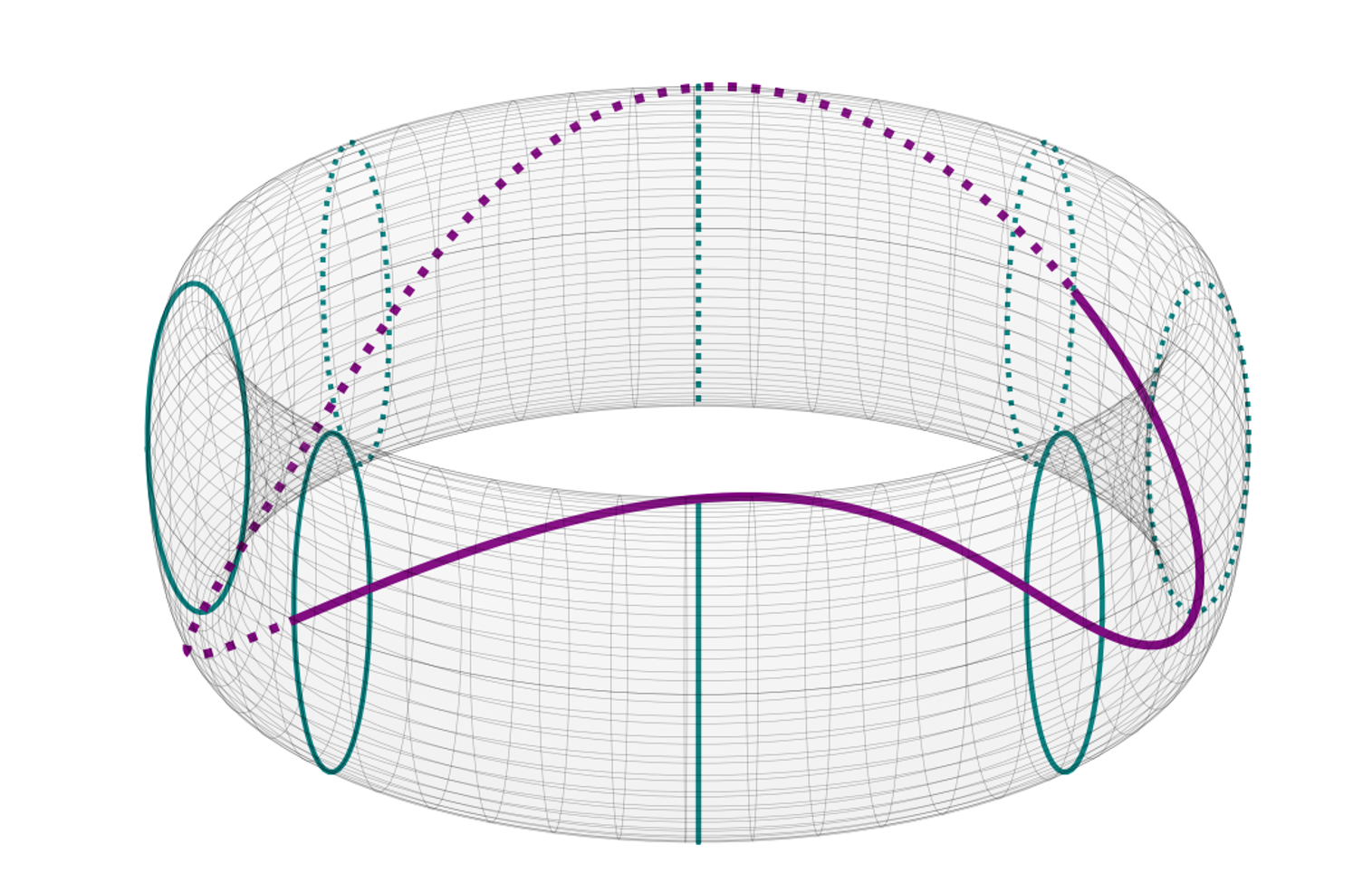}
    \caption{A Wilson loop on a foliated Riemann surface of genus 1 which is obtained from gluing a sleeve to itself. The leaves of the foliation are denoted as teal circles.}
    \label{fig:enter-label1}
\end{figure}
The transverse component cannot be gauged away as discussed in \secref{sec:Moduli}. As a sanity check, one can see that the infinitesimal gauge transformations do not act on $A_\theta$ in the abelian case.  Consequently, we obtain the isomorphism
\begin{equation}
\mathcal{M}\left(\mathbb{T} P^1, G\right) \cong \operatorname{Hom}\left(\pi_1\left(S^1\right), G\right) / G \cong \operatorname{Hom}\left(\\\mathbb{Z}, G\right) / G,
\end{equation}
where the $S^1$ is the tropical circle that foliated the complex manifold $\mathbb{TP}^1$. Although the moduli space of tropical flat connections is sensitive only to the leaves of the foliation, the curve is still able to wiggle around in the transverse directions. Notice that this is in contrast to the standard BF theory where the moduli space of flat connections would collapse to a single point because $\mathbb{CP}^1$ is simply connected. The Maslov dequantization limit of the path integral instead picks out winding numbers associated to the leaves of the foliation along the $\theta$ direction.

In order to extend this construction to higher genus, we first begin with the exceptional case of the torus which can be obtained by gluing the sleeve with itself to form a foliated torus. Since tropical connections are insensitive to radial parallel transport, the radial matching condition is trivially satisfied, and consequently, we only get contributions from the angular holonomy. The moduli space of the tropicalized flat connection can only obtain moduli from the orientation of $A_\theta$ and constraints from the way in which the $A_\theta$ associated to different sleeves must be glued together. The gluing must be done in such a way that the orientation of $A_\theta$ is preserved; for the case of a foliated torus, this is trivially implemented, which once again results in the moduli space of the tropicalized flat connections being $\operatorname{rank}(\mathfrak{g})$ i.e., 
\begin{equation}
    \operatorname{dim}\mathcal{M}(T^2,G)=\operatorname{rank}(\mathfrak{g}) ,
\end{equation}
where $T^2$ is the foliated torus. Notice that here, we have a very explicit departure from BF theory, where the moduli space of flat connections for a torus would give $2 \operatorname{rank}(\mathfrak{g})$. In order to extend this to a foliated compact Riemann surface of genus $g$, we need to investigate how the moduli space of tropicalized flat connections behaves on a foliated pair of pants. Unlike the foliated torus, the foliated pair of pants will have additional constraints due to the juncture where the foliations come together.

For each foliated cylinder the only free datum is the holonomy along the leaves of the foliation which contributes $\operatorname{rank}(\mathfrak{g})$. When you form a pair of pants by gluing three sleeves together, you would naively start with 3 $\operatorname{rank}(\mathfrak{g})$ parameters, one for each boundary. However, the gluing at the juncture forces the three boundary holonomies to multiply to the identity since a loop around the entire pair of pants is contractible. In the maximal torus, this gives a constraint of $\operatorname{rank}(\mathfrak{g})$ equations. That is, you must have

$$
g_1 g_2 g_3=1,
$$
with each holonomy $g_i$  specified by $\operatorname{rank}(\mathfrak{g})$ parameters up to conjugacy.  Thus, the matching condition reduces the total by $\operatorname{rank}(\mathfrak{g})$  dimensions, thus the moduli space of tropicalized flat connections on a foliated pair of pants $\mathcal{P}$ is
\begin{equation}
    \operatorname{dim}\mathcal{M}(\mathcal{P},G)=2\operatorname{rank}(\mathfrak{g}) .
\end{equation}
We note that in gluing the three sleeves to form a pair of pants, you can always set two of the sleeves to have the same orientation on the juncture but the third one will have an orientation that is opposite to at least one of the other two sleeves. This tells us that the gluing rules for anisotropic topological field theories has a refinement, in which we have to label the orientation of the foliation, when constructing higher genera foliated Riemann surfaces. We interpret this as a refinement of the usual Atiyah-Segal axioms of topological field theory in the same spirit as was suggested in \cite{Albrychiewicz:2023ngk}.

For the case of $g\geq 2$,  we want to construct a foliated Riemann surface $\Sigma_g$ of genus g through gluing together $2g-2$ foliated pair of pants with the appropriate orientations. The number of independent gluing circles is, therefore, half the original number of boundary circles provided by the $2g-2$ foliated pair of pants, which is equal to $3(2g-2)$. Each of the gluings comes with a matching condition that removes $\operatorname{rank}(\mathfrak{g})$ parameters and therefore we obtain
\begin{align}
    \dim \CM (\Sigma_g, G)=(g-1)\text{rank}(\mathfrak{g}).
\end{align}
With the moduli space dimension now known, at least away from reducible connections, one is able to readily compute the path integral through standard BRST cohomological methods. We leave this as an open question for future work.

\section{Examples of TBF Theories }
\label{sec:ExamplesTBF}
In \secref{sec:Moduli}, we saw that we are able to calculate the dimension of the moduli space of tropical flat connections through a twisted cohomological argument and show that its dimension on a sleeve is given generically by
\begin{equation}
\operatorname{dim} \mathcal{M}\left(\mathbb{TP}^1, G\right)=\operatorname{rank}(\mathfrak{g}).
\end{equation}
We expect this formula to generically hold away from singular connections where the dimension of the moduli space can jump. We now verify this statement, away from singular points, by using the isomorphism
\begin{equation}
\mathcal{M}\left(\mathbb{T} P^1, G\right) \cong \operatorname{Hom}\left(\pi_1\left(S^1\right), G\right) / G \cong \operatorname{Hom}(\mathbb{Z}, G) / G.
\end{equation}
A homomorphism $\mathbb{Z}\rightarrow G$, is determined by choosing a group element $g\in G$, but since $\mathbb{Z}$ is generated by a single element, the conjugation action of $G$ then acts on this element by $h \cdot g=h g h^{-1}$, thus this space classifies conjugacy classes in $G$. If we restrict to the case of compact Lie groups like SU$(N)$ or SO$(N)$, then the set of conjugacy classes of $G$ is parametrized by its maximal torus $T\subset G$ modulo the action of the Weyl group $W(G)$, thus our moduli space of tropicalized flat connections is given by
\begin{equation}
\mathcal{M}\left(\mathbb{T} P^1, G\right) \cong T / W(G).
\end{equation}
We will compute some simple, nontrivial examples of the above in the next subsections.

\subsection{SU(2) Case}
For the first nontrivial case of SU$(2)$, we can explicitly compute $\operatorname{Hom}(\mathbb{Z}, \text{SU}(2))$ by simply choosing the image of the generator of $\mathbb{Z}$ and thus, we obtain $\operatorname{Hom}(\mathbb{Z}, \text{SU}(2)) \cong \text{SU}(2)$. The dimension of $\text{SU}(2)$ is $3$. Now, we compute the quotient where SU$(2)$ acts on itself by conjugation. Since SU$(2)$ is a compact Lie group, its conjugacy classes are parametrized by the maximal torus modulo the Weyl group. A maximal torus in SU$(2)$ can be given by the subgroup
\begin{equation}
T=\left\{\left.\left[\begin{array}{cc}
e^{i \theta} & 0 \\
0 & e^{-i \theta}
\end{array}\right] \right\rvert\, \theta \in[0, \pi]\right\} \cong U(1) .
\end{equation}
The Weyl group of SU$(2)$ is $W(\text{SU}(2)) \cong \mathbb{Z}_2$, which acts by $\theta \mapsto-\theta$. Altogether, we have that
\begin{equation}
T / W(\text{SU}(2)) \cong[0, \pi] \cong \mathbb{R}^1,
\end{equation}
hence, 
\begin{equation}
\operatorname{dim} \mathcal{M}\left(\mathbb{T} P^1, \text{SU}(2)\right)=1.
\end{equation}
This matches the result that we obtained from the twisted cohomology complex which states $\operatorname{dim} \mathcal{M}\left(\mathbb{T P}^1, \text{SU}(2)\right)=\operatorname{rank}(\mathfrak{su}(2))=1$.

\subsection{SU$(N)$ Case}
Given these explicit examples, we show the equivalence for general $N$. In fact, it is well known that  $\text{Hom}(\text{SU}(N), \mathbb{Z})$ is isomorphic to the maximal torus of SU$(N)$ i.e. $\text{Hom}(\text{SU}(N), \mathbb{Z})\cong \text{U}(1)^{N-1}$.  The quotient by conjugation leaves the dimension of the torus unchanged so we conclude
\begin{align}
    \operatorname{dim} \mathcal{M}\left(\mathbb{T} P^1, \text{SU}(N)\right)=N-1,
\end{align}
which, along with the fact that the $\operatorname{rank}(\text{SU}(N))=N-1$ , coincides with twisted cohomology methods which gives
\begin{align}
  \operatorname{dim} \mathcal{M}\left(\mathbb{T P}^1, \text{SU}(N)\right)=\operatorname{rank}(\mathfrak{su}(N)) =N-1.
\end{align}

\subsection{U(1) Case: Singular Tropical Connections}
We can be very explicit about the case where the gauge group is U$(1)$. Here, the moduli space of tropicalized flat connections is isomorphic to
\begin{equation}
\mathcal{M}\left(\mathbb{T} P^1, \text{U}(1)\right)  \cong \operatorname{Hom}(\mathbb{Z}, \text{U}(1)) / \text{U}(1) \cong \text{U}(1)/\text{U}(1),
\end{equation}
where the second isomorphism is given by the fact that a homomorphism from $\mathbb{Z}$ to U$(1)$ is completely determined by the image of the generator $1\in \mathbb{Z}$ under $\phi(n)=z^n$  for some $z \in \text{U}(1)$. Quotienting this out gives us a zero-dimensional space and thus we get
\begin{equation}
\operatorname{dim} \mathcal{M}\left(\mathbb{T} P^1, U(1)\right)=0,
\end{equation}
however, this is in direct contradiction to the fact that the rank of $U(1)$ is 1. This is suggestive of the fact that in the tropical case, we still have a notion of reducible/singular connection, where the moduli space dimension can bubble/jump. The discrepancy arises from the fact that in the calculation for the dimension of the moduli space given by the twisted cohomology of the deformation complex, we assumed that the background tropical flat connection was irreducible. If we have a reducible tropical connection, we may have a larger stabilizer subgroup which would reduce the dimension of the cohomology even further. 

\section{A Glossary of Tropical-Topological (Tropological) Field Theories}
\label{sec:Glossary}

We have now seen that we are able to construct a consistent anisotropic topological field theory out of the tropical limit of conventional isotropic topological field theories. One of the most interesting outcomes of this work is that the construction of 2-dimensional TBF theory naturally opens the door to a wealth of new tropical topological field theories. A natural next step is to extend the tropicalization prescription to other Schwartz-type theories. In this section, we will discuss how this construction allows us to easily construct tropical analogs of 2-dimensional topological Yang-Mills theory (TrYM). We will then extend the construction to 3-dimensional TBF theories which then give us a prescription for formulating 3-dimensional tropical Chern-Simons theory (TCS). If we place Chern-Simons theory on a manifold with boundary, then one can also see how this construction extrapolates to tropical Wess-Zumino models which could have a potential relation to topological analogs of fractons. It is conjectured that the tropical Wess-Zumino Witten model (TWZW) can be constructed as an anisotropic generalization of topological field theory as suggested by \cite{Albrychiewicz:2023ngk}.

\subsection{2D Tropological Yang Mills Theory (TrYM) }
It is well known that 2D Yang-Mills theory \cite{Migdal:1975zg} can be related to the 2D BF theory. We can explicitly show this by starting with the Yang-Mills functional on a Riemann surface $\hat{\Sigma}$
\begin{equation}
S_{Y M}=\frac{1}{4 g^2} \int_{\hat{\Sigma}} \operatorname{Tr}(F \wedge * F).
\end{equation}
This can be written in the first order formalism by introducing an adjoint valued B field and a coupling constant $g$ as
\begin{equation}
S_{YM}=\frac{1}{2}\int_{\hat{\Sigma}} \operatorname{Tr}(B \wedge F)-\frac{g^2}{4} \int_{\hat{\Sigma}} \operatorname{Tr}\left(B\wedge \star_J B\right).
\end{equation}
One can see that the resulting action has a BF like structure for nonzero coupling constant and in the limit of zero coupling, one obtains precisely the 2D BF theory. Here, we utilized a modified Hodge star known as the Jordan star as explained in \cite{}.  A detailed review can be found in \cite{Cordes:1994fc}. We can use the same prescription for tropicalizing this theory as we did for the case of BF theory, however, given that $B\rightarrow \frac{B}{\hbar}$, we must also scale our Yang-Mills coupling by $g\rightarrow\hbar g$. In the tropical limit, we have, what we call \textit{2D Tropological Yang Mills} in the first order formalism on a foliated Riemann surface $\Sigma$ 
\begin{equation}
S_{Tr YM}=\int_{\Sigma} d r d \theta \hspace{0.1cm}\operatorname{Tr}\left[\frac{1}{2}T\left(-\partial_\theta A_r\right)+\frac{1}{2}B\left(\partial_r A_\theta+\left[A_r, A_\theta\right]\right)\right]-\frac{g^2}{4} \int_{\Sigma} \operatorname{Tr}\left(B^2\right).
\end{equation}
We abbreviate this as \textit{TrYM} theory in order to not have any conflicts with topological Yang Mills theory which some authors denote as TYM. 

Integrating out the auxiliary \(B\)-field then yields an effective theory whose action involves a squared transverse curvature term, very much in the spirit of standard YM theories but now with explicit anisotropic features that encode the underlying foliation. The action is
\begin{equation}
S_{TrYM}=\int_{\Sigma} d r d \theta \operatorname{Tr}\left[\frac{1}{2}T (-\partial_\theta A_r)+\frac{1}{2 g^2}\left(\partial_r A_\theta+\left[A_r, A_\theta\right]\right)^2\right].
\end{equation}
This action describes 2D tropological Yang-Mills theory in the second order formalism. The equations of motion are given by 
\begin{equation}
\begin{aligned}
&\partial_\theta A_r=0 ,\\
&\frac{1}{2}\partial_\theta T+\frac{1}{g^2}\left[A_\theta,\left(\partial_r A_\theta+\left[A_r, A_\theta\right]\right)\right]=0,\\
&\partial_r\left(\partial_r A_\theta+\left[A_r, A_\theta\right]\right)+\left[A_r, \partial_r A_\theta+\left[A_r, A_\theta\right]\right]=0.
\end{aligned}
\end{equation}
Just like TBF theory, tropical Yang Mills theory is also invariant under a restricted class of gauge transformations, namely foliation preserving gauge transformations. 

One can now investigate a rich array of questions emerging from this tropicalization of 2D topological Yang-Mills theory. For instance, how does the tropical scaling of the coupling constant and the rescaling of the $B$-field influence the spectrum compared to the conventional Yang-Mills and BF theories? How is the underlying representation theory changed? How are the fusion rules and conformal blocks modified? What new insights can be gleaned from the explicit anisotropic features introduced by the foliation, and how do these features manifest in physical observables such as the partition function and Wilson loop correlation functions? Can these observables be probed by an anisotropic generalization of equivariant localization now that the localization locus is modified? Can one obtain this theory in terms of a lattice formulation? Does it have a large $N$-expansion and a related tropical string theory associated to it? Lastly, one can include an analysis of the classical solutions and attempt to find solitionic configurations for TrYM. We leave these as open questions which should have an conclusive answer in a finite amount of time.

\subsection{Higher Dimensional TBF Theory}
In this paper, we have focused on 2D BF theory but it is well known that BF theory in higher dimensions provides an even richer central topological field theory since it acts as a prototypical example of a topological field theory with higher form symmetries \cite{Gaiotto_2015}. Tropicalizing these theories might provide fresh perspectives on higher form symmetries in anisotropic topological field theory. It is expected that the associated forms now have an additional projectability condition in order to be compatible with the foliation. Furthermore, 3-dimensional BF theory has an obvious connection with 3D Chern-Simons theory. In fact, it can be demonstrated that for 3D BF theory can be cast as a Chern-Simons theory with a noncompact gauge group that acts naturally on the tangent bundle of the gauge group of the BF theory \cite{Witten:1988hf,Blau:2022krm}. Lastly, it has also been demonstrated that higher dimensional BF theories can be used for descriptions of gravitational systems. Most notably, 3D Einstein gravity is known to not have any propagating degrees of freedom and can be related to a BF theory with an SO$(1,2)$ gauge group for Lorentzian signature \cite{freidel1999bfdescriptionhigherdimensionalgravity} and an SO$(3)$ gauge group for Euclidean signature. Likewise, higher-dimensional constrained BF theories can be used to describe higher-dimensional Einstein gravities. 

With all these motivations in mind, we now provide some of the basic details for this construction on a simple product geometry by placing BF theory on a 3-dimensional manifold $M_3 =\mathbb{CP}^1 \times \mathbb{R}$, the Lagrangian action is
\begin{equation}
S_{\mathrm{BF}}=\int_{M_3} \operatorname{Tr}(B \wedge F(A)).
\end{equation}
In this scenario, the $B$ field of BF theory is an adjoint-valued 1-form field and we have an additional reducible topological shift symmetry that introduces second generation ghosts. We proceed as in the two-dimensional case and take the Maslov dequantization limit of the $\mathbb{CP}^1$ while leaving the extra $\mathbb{R}$ alone. We parametrize the coordinate on the $\mathbb{R}$ via $\phi$.  Expanding the terms in the action, we get
\begin{equation}
S^{3D}_{B F}=\int_{\mathbb{CP}^1 \times \mathbb{R}} d z d \bar{z} d \phi \operatorname{Tr}\left(B_z F_{\bar{z} \phi}+B_{\bar{z}} F_{\phi z}+B_\phi F_{z \bar{z}}\right) .
\end{equation}
The additional complication for 3D BF theory is that the holomorphic components and antiholomorphic components of $B$ must transform under the Maslov dequantization procedure as
\begin{equation}
\begin{aligned}
& B_z=e^{-r / \hbar-i\theta}\left(\frac{\hbar}{2}B_r-\frac{i}{2} B_\theta\right), \\
& B_{\bar{z}}=e^{-r / \hbar+i \theta}\left(\frac{\hbar}{2} B_r+\frac{i}{2} B_\theta\right).
\end{aligned}
\end{equation}
In order to make contact with our original prescription for 2D TBF, we know that we must perform the same transformations on $A_\theta$ and $B_\phi$ as before i.e., the anisotropic scaling limit is
\begin{equation}
\begin{array}{ll}
A_r \rightarrow  A_r, & B_r \rightarrow B_r, \\
A_\theta \rightarrow \hbar A_\theta, & B_\theta \rightarrow B_\theta, \\
A_\phi \rightarrow A_\phi, & B_\phi \rightarrow \frac{B_\phi}{\hbar}.
\end{array}
\end{equation}
A short exercise in algebra then demonstrates that the tropical 3D TBF theory is given by the Lagrangian action
\begin{equation}
\begin{aligned}
S^{3D}_{TBF}=\int_{\mathbb{TP}^1\times \mathbb{R}} d r d \theta d \phi \operatorname{Tr}[ & B_r\left(\partial_\theta A_\phi\right)-B_\theta\left(\partial_r A_\phi-\partial_\phi A_r+\left[A_r, A_\phi\right]\right) \\
& \left.+B_\phi\left(\partial_r A_\theta+\left[A_r, A_\theta\right]\right)+T_\phi\left(-\partial_\theta A_r\right)\right].
\end{aligned}
\end{equation}
The equations of motions of $B$ and $T$ give the following flatness and projectability constraints
\begin{equation}
\begin{aligned}
& F_{r \phi}=\partial_r A_\phi-\partial_\phi A_r+\left[A_r, A_\phi\right] =0,\\
& f_{r \theta}=\partial_r A_\theta+\left[A_r, A_\theta\right]=0, \\
& \partial_\theta A_r=0, \\
& \partial_\theta A_\phi=0.
\end{aligned}
\end{equation}
Along with the covariant constancy conditions coming from the equations of motion for $A$
\begin{equation}
\begin{aligned}
& \partial_\theta T_\phi-\partial_\phi B_\theta-\left[A_\phi, B_\theta\right]+\left[A_\theta, B_\phi\right]=0, \\
& \partial_r B_\phi+\left[A_r, B_\phi\right]=0, \\
& \partial_r B_\theta-\partial_\theta B_r+\left[A_r, B_\theta\right]=0.
\end{aligned}
\end{equation}

The structure of 3D TBF is somewhat expected from the Maslov dequantization limit and the discussion in \appref{app:Ineq}. Here, the components of the tropicalized connection $A$ that are not associated to the $\theta$-direction that runs along the leaves of the foliation, namely $A_r$ and $A_\phi$ both occur with a constraint, which says, that they must be projectable onto the foliation. Notice that we recover a standard antisymmetric curvature tensor $F$ along the $(r,\phi)$ direction, but a deformed "transverse" curvature tensor $f$ for the $(r,\theta)$. We can see that at the level of the equations of motion of $A^\theta$ tropicalization does not affect the directions that are transverse to the entire foliation on $\mathbb{TP}^1$ with the caveat that the foliation structure imposes some minimal consistency conditions, which states that the fields $A_r, A_\phi$ must be projectable. 

In the case of 2D TBF theory, it was clear that all the components of the connection $A$ had been affected by tropicalization. In higher-dimensional TBF theories, it can be expected that tropicalization deforms the curvature tensors along the direction that is tropicalized while the other directions retain their structure up to projectability conditions. Despite this, there is nothing stopping us from preforming multiple tropicalizations to induce several overlapping foliations. As was discussed in \cite{Albrychiewicz:2023ngk}, for the case of 3D geometries, one can consider a higher rank Jordan structure of the form
\begin{equation}
J=\left(\begin{array}{lll}
0 & 1 & 0 \\
0 & 0 & 1 \\
0 & 0 & 0
\end{array}\right) .
\end{equation}
This induces the structure of a nested double foliation on the underlying geometry. This simple example shows that despite tropicalization generically being for even-dimensional complex spaces, one is able to naturally extend the Maslov dequantization procedure to odd-dimensions and go beyond the realm of tropical geometry. It is an interesting open question of what sort of mathematical structure this sort of Jordan structure can arise from. For the case of even higher dimensional manifolds, some preliminary results were given in\cite{Albrychiewicz:2025afk}.

\subsection{Tropical Chern-Simons Theory (TCS)}
Now that it is clear how to generalize TBF theory to higher dimensions, we can leverage the construction by trying to see how it extrapolates to more complicated topological field theories. One of the most prominent 3D topological field theories that can be considered is Chern-Simons theory. In order to see why one would like to tropicalize Chern-Simons, we will begin with some preliminary motivations.

Although Chern-Simons theory first appeared in strictly mathematical considerations \cite{Chern1974CharacteristicFA}, it was soon shown that it can be interpreted as a topological quantum field theory which naturally appears in many different contexts throughout quantum field theory and string theory \cite{Witten:1988hf, witten2003chernsimonsgaugetheorystring,Mari_o_2005,Bershadsky:1993cx}, such as a tool in the classification of topological matter \cite{Chiu:2015mfr}, as an explicit example in supersymmetric path integral localization \cite{Kapustin:2009kz,Mari_o_2011} and as a theory which many other quantum field theories are dual to \cite{dimofte2011chernsimonstheorysduality}.

Recently, it was discovered that tropical analogs of Lagrangian and coisotropic branes \cite{Albrychiewicz:2024tqe,Albrychiewicz:2025hzt} can be constructed through worldsheet methods. It would be interesting to know if the tropicalized Chern-Simons theories that are presented below appear as the natural worldvolume theory for the tropical A-branes similar in the spirit of \cite{witten2003chernsimonsgaugetheorystring} where, for instance, it was shown that when you wrap A-branes along Lagrangian submanifolds, the low-energy effective field theory on their worldvolume is a 3D Chern-Simons theory. 

Chern-Simons theory is not only relevant for theoretical considerations, it has significantly influenced quantum information science and quantum computing. Its inherent topological features have inspired novel approaches to fault-tolerant quantum computation through topologically protected states such as anyons \cite{Nayak:2008zza,AROVAS1985117}. In condensed matter physics, Chern-Simons theory is often a low-energy effective description for many physical phenomena, for example, it becomes relevant in the description of the emergence of topological insulator edge states in the context of the fractional quantum Hall effect \cite{Hasan:2010xy,Qi:2010qag}.

It is clear that finding a tropical generalization of Chern-Simons theory could potentially lead to interesting anisotropic generalizations of the aforementioned results. We begin with the Lagrangian action for 3D Chern-Simons theory on the geometry $\mathbb{CP}^1 \times \mathbb{R}$
\begin{equation}
S_{C S}[A]= \frac{k}{4\pi}\int_{\mathbb{CP}^1 \times \mathbb{R}} \operatorname{Tr}\left(A \wedge d A+\frac{2}{3} A \wedge A \wedge A\right).
\end{equation}
We will see that in order to implement tropicalization for Chern-Simons theory, it is best to first focus on the kinetic term and then later focus on the interaction term. Thus, we will begin by first focusing on abelian Chern-Simons theory. If one tries to reapply our prescription for 3D BF theory by identifying $B_\phi$ with $A_\phi$ i.e.
\begin{equation}
\begin{aligned}
& A_r \rightarrow A_r ,\\
& A_\theta \rightarrow \hbar A_\theta, \\
& A_\phi \rightarrow \frac{1}{\hbar} A_\phi.
\end{aligned}
\end{equation}
One will notice that the scaling between terms in the term $A \wedge d A$ is mixed since $A_\phi$ enters in each of the terms and the limit cannot be properly taken. Instead, the prescription must be refined and one must allow the scaling that was originally given by deforming $A_\phi$ to instead be implemented by renormalizing the Chern-Simons level $k$. The prescription for obtaining tropical Chern-Simons theory is then
\begin{equation}
\begin{aligned}
& A_r \rightarrow A_r, \\
& A_\theta \rightarrow \hbar A_\theta, \\
& A_\phi \rightarrow A_\phi, \\
& k \rightarrow \frac{k}{\hbar}.
\end{aligned}
\end{equation}
Performing this field redefinition along with the Maslov dequantization on the underlying geometry of the abelian 3D Chern-Simons functional then yields the abelian 3D tropical Chern-Simons (TCS) action 
\begin{equation}
\begin{aligned}
S_{TCS}=\frac{k}{4\pi}\int_{\mathbb{TP}^1\times \mathbb{R}} dr d\theta d\phi & {\left[A_\phi \partial_r A_\theta-A_r \partial_\phi A_\theta+T_r\left(\partial_\theta A_\phi\right)+T_\phi\left(-\partial_\theta A_r\right)\right.} \\
& \left.+A_\theta\left(\partial_\phi A_r-\partial_r A_\phi\right)\right],
\end{aligned}
\end{equation}
here we have performed two Hubbard-Stratonovich transformations. 

The equations of motion for the components of $T$ enforce the projectability of the fields $A_r$ and $A_\phi$ which do not run along the leaves of the foliation
\begin{align}
\label{eqn:CSProjCond1}
    \partial_\theta A_\phi&=0, \\
\label{eqn:CSProjCond2}    
    \partial_\theta A_r&=0.
\end{align}
The equations of motion for the components of $A$ then give
\begin{align} \nn
    f_{r\theta}&=\partial_r A_\theta=\frac{1}{2}\partial_\theta T_r ,\\ \nn
    f_{\theta\phi}&=-\partial_\phi A_\theta=-\frac{1}{2}\partial_\theta T_\phi, \\ \nn
    F_{r \phi}&=\partial_r A_\phi-\partial_\phi A_r=0. 
\end{align}
The structure of tropical Chern-Simons theory is similar to the TBF theory in the sense that along the $(r,\theta)$ and $(\phi,\theta)$-directions, we no longer have the usual flatness condition, which is expressed as an antisymmetric 2-tensor but instead, it is deformed into a non-symmetric object $f_{r\theta}$, $f_{\phi\theta}$ respectively. Along the $(r,\phi)$-directions, we have a standard flatness condition. In this case, the $(r,\phi)$ components of the connection $A$ is insensitive to the leaves of the foliation and the additional $\mathbb{R}$ acts as an overall global factor. We will see that in the non-abelian case, this is no longer true because the underlying Lie-algebra couples all directions. In order to generalize this to the nonabelian case, we need to investigate what happens to the interaction term under our prescription for tropicalizing Chern-Simons theory
\begin{equation}
\frac{k}{4\pi}\int_{\mathbb{CP}^1 \times \mathbb{R}} \operatorname{Tr}\left(\frac{2}{3} A \wedge A \wedge A\right).
\end{equation}
Besides the fact that the complex projective space becomes tropical projective space, this term remains invariant under tropicalization and thus the full nonabelian tropical Chern-Simons theory is given by the following action
\begin{equation}
\begin{aligned}
\label{eqn:TropNonAbCS}
S_{TCS}=\frac{k}{4\pi}\int_{\mathbb{TP}^1\times \mathbb{R}} d r d\theta d\phi \operatorname{Tr} & {\left[A_\phi\left(\partial_r A_\theta+ \frac{2}{3}\left[A_r, A_\theta\right]\right)+T_\phi\left(-\partial_\theta A_r\right)\right.} \\
& +A_r\left(-\partial_\phi A_\theta+\frac{2}{3}\left[A_\theta, A_\phi\right]\right)+T_r\left(\partial_\theta A_\phi\right) \\
& \left.+A_\theta\left(\partial_\phi A_r-\partial_r A_\phi+\frac{2}{3}\left[A_\phi, A_r\right]\right)\right].
\end{aligned}
\end{equation}

The non-abelian theory retains the projectability conditions \eqref{eqn:CSProjCond1}-\eqref{eqn:CSProjCond2} but the equations of motion for the components of the connection $A$ get modified to
\begin{align} \nn
    f_{r \theta}&=\partial_r A_\theta+\left[A_r, A_\theta\right]=\frac{1}{2}\partial_\theta T_r, \\ \nn
    f_{\theta \phi}&=-\partial_\phi A_\theta+\left[A_\theta, A_\phi\right]=-\frac{1}{2}\partial_\theta T_\phi, \\ \nn
    F_{\phi r}&=\partial_\phi A_r-\partial_r A_\phi+\left[A_\phi, A_r\right]=0. 
\end{align}
The infinitesimal foliation-preserving gauge transformations for TCS are now
\begin{align}
    \label{eqn:AiGauge}
    \delta A_i&=\partial_r\lambda(r,\phi)+[A_i, \lambda(r,\phi)], \quad i=r,\phi, \\ \label{eqn:AThetaGauge}
    \delta A_\theta&=[A_\theta, \lambda(r,\phi)], \\
    \delta T_i&=[T_i, \lambda(r,\phi)], \quad i=r,\phi. 
\end{align}
In order to obtain this action, we were essentially forced to renormalize the Chern-Simons level by allowing $k\rightarrow \frac{k}{\hbar}$. In taking the tropical limit, where $\hbar\rightarrow 0$, this is suggestive that the tropical Chern-Simons theory is related to the usual Chern-Simons theory in the infinite level limit, implying that the tropicalized version should be one-loop exact regardless of tropical level $k$. One might expect this sort of behavior to occur generically for all field theories obtained through tropicalization as was shown in \cite{Albrychiewicz:2025rkg}, where it was found that tropicalizing the Dirac equation introduces a new purely fermionic topological symmetry that is reminiscent of supersymmetric localization.

It is well known that the Chern-Simons functional on a manifold with boundary acquires a boundary term that allows one to introduce a chiral Wess-Zumino-Witten model related to the underlying gauge symmetry of the Chern-Simons theory. We will demonstrate that the same phenomena occurs for the TCS theory for a simple geometry. Consequently, we will put the CS theory on a manifold with a boundary at $\phi=0$:  $\mathbb{CP}^1 \times I_{[0,\infty)}$ and take a tropical limit. We will choose the boundary conditions at $\phi \rightarrow\infty$ such that the field configurations vanish.

If we perform a foliation-preserving gauge transformation on the action \eqref{eqn:TropNonAbCS}, we find that we have a leftover total derivative term along the $\phi$-direction
\begin{align}
    \frac{k}{4\pi}\int_{\mathbb{TP}^1 \times I_{[0,\infty)}} dr d\theta d\phi \; \tr[\partial_\phi(A_\theta \delta A_r-A_r\delta A_\theta)].
\end{align}
At the boundary, the leftover gauge transformations can be used to transform the fields $A_r$ and $A_\phi$ into pure gauge, however the $A_\theta$ cannot be completely gauged away, instead, it resides in the Cartan subalgebra of the Lie-algebra $\mathfrak{g}$. From \eqref{eqn:AiGauge} and \eqref{eqn:AThetaGauge}, we have
\begin{align}
    \delta A_r&=g^{-1}\partial_r g, \\
    \delta A_\theta&=g^{-1}Hg,  
\end{align}
where $H$ is in Cartan subalgebra and $g$ is a map of the form $g: \mathbb{TP}^1\rightarrow G $. Therefore, the classical tropical WZW (TWZW) term is of the form
\begin{align}
  S_{TWZW}=  \frac{k}{4\pi} \int_{\mathbb{TP}^1} dr d\theta \tr[g^{-1}H\partial_r g].
\end{align}

One of the most useful properties of WZW models comes from their conformal invariance. It is known that in order to construct a quantum mechanically consistent WZW model, one needs to first check whether or not this conformal symmetry is violated by any anomalies. In order to do so, the beta function must be computed and the appropriate counter-term must be added to preserve quantum conformal invariance. For the case of the TWZW, it is not clear whether or not this model admits a generalization of conformal invariance for two reasons, the first is that unlike the WZW model, the TWZW seems to be a first order derivative theory similar to the bosonic $\beta\gamma$ ghost systems that appear in the superconformal gauge fixing of the RNS superstring \cite{DiFrancesco:1997nk}. However, contrary to ghost systems that can carry a Kac-Moody Lie-algebra symmetry, the fields of TWZW are Lie group-valued. The second is that it is not clear whether or not there is a proper substitute for the statement that the stress energy tensor must be traceless in the tropical case. Investigating these questions would require a more detailed analysis but could lead to an explicit example of anisotropic conformal field theory.

\section{Conclusion and Discussions }
\label{sec:Conclusions}

In this paper, we have introduced and analyzed a tropicalization of the 2D BF theory, which we called the TBF theory for short. Our construction provides an anisotropic generalization of conventional BF theory that naturally incorporates a foliated complex geometry structure which represents the tropical projective space \(\mathbb{TP}^1\). In the tropical limit, the usual complex structure on \(\mathbb{CP}^1\) is deformed into a nilpotent Jordan structure whose associated foliation plays a central role in defining the foliation-preserving gauge symmetries of the theory.  

We showed that the path integral localized onto a moduli space of tropicalized flat connections and explicitly showed that the TBF theory retains nontrivial dynamics because the tropicalized manifold \(\mathbb{TP}^1\) is now interpreted as a foliated complex geometry whose leaves, which are topologically \(S^1\), support a nontrivial leaf-wise fundamental group. As a consequence, the moduli space of tropicalized flat connections on a sleeve, 
\[
\mathcal{M}\bigl(\mathbb{TP}^1, G\bigr) \cong \operatorname{Hom}(\mathbb{Z}, G)/G,
\]
remains finite-dimensional even for spaces that would be topologically trivial in the standard BF framework. We checked our analysis using both holonomy considerations and the twisted cohomology of the deformation complex to obtain the general result on a sleeve,
\[
\dim \mathcal{M}\bigl(\mathbb{TP}^1, G\bigr) = \operatorname{rank}(\mathfrak{g}),
\]
which we explicitly verified for the case \(G=\text{SU}(N)\). These computations serve as a robust consistency check for our tropical construction. 

After having done the analysis on the sleeve, we showed that we can glue multiple sleeves together as long as we take into account the matching conditions that appear when the foliations meet at a juncture. We found that for a generic foliated Riemann surface of genus $\Sigma_g$, with $g \geq 2$, this gives
\begin{equation}
\operatorname{dim} \mathcal{M}\left(\Sigma_g, G\right)=(g-1) \operatorname{rank}(\mathfrak{g}).
\end{equation}
Having shown that 2D TBF can be seen as a consistent anisotropic topological field theory, we then extended the prescription to several other interesting TQFTs such as 2D topological Yang-Mills, higher-dimensional BF theories and Chern-Simons theory. 

Beyond the TBF theory, there are still many questions that can be investigated in the context of anisotropic topological field theories. Although, we had constructed the Lagrangian for 2D tropical Yang-Mills theory, it is clear that there should be higher-dimensional analogs of tropical Yang-Mills that can be obtained through similar methods. One of the most interesting questions that one could ask is how tropicalization would affect 4D Yang-Mills instantons. In fact, one might expect this is precisely where tropical geometry could provide new insights into 4-dimensional topology by generalizing Mikhalkin's result \cite{mikhalkin} that tropicalization preserves the Gromov-Witten invariants to the Donaldson invariants. We conjecture that such a connection should exist given the fact that the generating functional of the Gromov-Witten invariants has an intricate connection to the generating functional of the Donaldson invariants through Seiberg-Witten theory \cite{taubes2005seiberg,moore1997integrationuplanedonaldsontheory}. While we have shown that one is able to get a result for the moduli space of tropicalized flat connections for the simple case of a foliated Riemann surface, the extension of tropicalization to higher-dimensions could introduce more complicated nested foliations, foliations with different co-dimensionality and singularities of foliations that could complicate the analysis. We leave this for future work.

We also conjecture that the TBF theory has natural connections to gravitational physics. For example, JT gravity has connections to the BF theory with gauge group $SL(2,\mathbb{R)}$ \cite{Turiaci:2024cad}. It is possible that a tropicalization of JT gravity could be linked to the TBF theory through the same arguments that are used in conventional JT gravity. If possible, this could naturally give rise to several other new directions such as novel anisotropic generalizations of holographic dualities. Such generalizations have been previously proposed under the name of Lifshitz holography \cite{Griffin_2013}.  

One could also consider the notion of tropicalized random matrix models \cite{saad2019jtgravitymatrixintegral} which arise from the supersymmetric localization of some of these theories. For example, it is well known that the matter sector of supersymmetric Chern-Simons theory admit a random matrix model description after the localization is explicitly computed. Likewise, JT gravity \cite{Jackiw:1984je, Teitelboim:1983ux}, which can be reformulated as BF theory, is known to admit random matrix model descriptions through its holographic dual.

We want to emphasize that our original motivation for investigating the tropical limits of these field theories is to explicitly construct simpler examples of the wedge region \cite{ssk, keq, neq} of the potential Schwinger-Keldysh extension of string theory worldsheet perturbation theory. The observations made in \cite{ssk, keq, neq} suggest that in order to probe the wedge region associated to the string worldsheet, one should consider a double analytic continuation as suggested in \cite{Witten:2013pra}. It has been seen in a series of papers by \cite{eberhardti, eberhardtii} that such a procedure will result in a tropical region that describes intermediate string states. Any intermediate string state is known to arise from asymptotic regions of the moduli space of punctures Riemann surfaces which can be equivalently be seen as a pinch of the worldsheet geometry. As a result, any pinching structures should be described by these tropical limits. Interestingly in \cite{DiUbaldo:2025cmg}, it was shown that if one performs a double analytic continuation as described above for two point correlators, this results in matrix models whose spectral correlators are neatly described by minimization operations which is the hallmark of tropical geometry. We conjecture that the correct notion of tropical matrix models should be of this form.

A striking feature of TBF theory is the appearance of subsystem symmetries associated with foliation-preserving diffeomorphisms. In contrast to the full diffeomorphism invariance of standard topological field theories, the restriction to foliation-preserving transformations means that gauge transformations act independently on each leaf of the foliation. This mirrors the structure of fracton models in condensed matter physics, where mobility restrictions arise due to subsystem symmetries. These observations suggest that the tropical limit of BF theory might serve as a fertile testing ground for exploring the interplay between topological field theories and the emergent physics of potentially topologically fracton phases. It also raises the possibility that a deeper categorization of gauge symmetries in tropicalized models may eventually lead to a new understanding of exotic phases of matter. On that note, if an explicit correspondence with fractonic field theories can be made, it would certainly be interesting to know what sort of lattice formulations can be proposed for TBF theory and the associated anisotropic topological field theories that were constructed in this paper.

\acknowledgments
This short paper was inspired by a private conversation with Ori Ganor on obtaining alternative compactifications of \cite{Albrychiewicz:2025uam}. We want to thank Petr Ho\v rava for insightful discussions, and also thank Viola Zixin Zhao for useful comments. Although, we have mentioned a extensive list of open questions above, this is in no way an exhaustive list, we encourage readers to provide email feedback on potential extensions. This work has been supported by the Leinweber Institute for Theoretical Physics.

\appendix 

\section{Nilpotency of the Deformation Complex}
\label{app:Nil}
In this short appendix, we will show that the deformation complex associated to the tangent space of the moduli space of tropicalized flat connections $\mathcal{M}\left(\mathbb{T P}^1, G\right)$ has a nilpotent differential. Recall that the deformation complex was given by
\begin{equation}
0 \rightarrow C^0 \xrightarrow{d_0} C^1 \rightarrow 0,
\end{equation}
where the space of gauge parameters is
\begin{equation}
C^0=\left\{\lambda(r) \in \Omega^0\left(\mathbb{T P}^1, \operatorname{ad} G\right) \mid \partial_\theta \lambda=0\right\},
\end{equation}
and the space of tropical connection deformations is
\begin{equation}
C^1=\left\{a=a_r(r) d r+a_\theta(r, \theta) d \theta \mid \partial_\theta a_r=0\right\}.
\end{equation}
We define the differential $d_0: C^0 \rightarrow C^1$ as 
\begin{equation}
d_0 \lambda=\left(\partial_r \lambda+\left[A_r, \lambda\right]\right) d r+\left[A_\theta, \lambda\right] d \theta,
\end{equation}
and
\begin{equation}
d_1 a \equiv \partial_r a_\theta+\left[A_r, a_\theta\right]+\left[a_r, A_\theta\right]=0.
\end{equation}
We want to show that $d_1 \circ d_0=0$.  Explicitly, we get that
\begin{equation}
d_1\left(d_0 \lambda\right)=\partial_r\left(\left[A_\theta, \lambda\right]\right)+\left[A_r,\left[A_\theta, \lambda\right]\right]+\left[\partial_r \lambda+\left[A_r, \lambda\right], A_\theta\right],
\end{equation}
which can be expanded into
\begin{equation}
d_1\left(d_0 \lambda\right)=\left[\partial_r A_\theta, \lambda\right]+\left[A_\theta, \partial_r \lambda\right]+\left[A_r,\left[A_\theta, \lambda\right]\right]+\left[\partial_r \lambda, A_\theta\right]+\left[\left[A_r, \lambda\right], A_\theta\right] .
\end{equation}
We can use the fact that Lie bracket is antisymmetric and that the Lie algebra still satisfies a Jacobi identity $\left[A_r,\left[A_\theta, \lambda\right]\right]+\left[\left[A_r, \lambda\right], A_\theta\right]=\left[\left[A_r, A_\theta\right], \lambda\right]$ to write
\begin{equation}
d_1\left(d_0 \lambda\right)=\left[\partial_r A_\theta, \lambda\right]+\left[\left[A_r, A_\theta\right], \lambda\right]=\left[\partial_r A_\theta+\left[A_r, A_\theta\right], \lambda\right] .
\end{equation}
However, we recall that we must satisfy the transverse flatness condition
\begin{equation}
\partial_r A_\theta+\left[A_r, A_\theta\right]=0 ,
\end{equation}
and so we get our result
\begin{equation}
d_1 \circ d_0=0.
\end{equation}
\section{Inequivalent Tropicalizations and Arctic Limits}
\label{app:Ineq}
We saw in \secref{sec:TropBF} that one is able to construct the two-dimensional TBF action by supplementing the Maslov dequantization $z=e^{\frac{r}{\hbar}+i \theta}, \bar{z}=e^{\frac{r}{\hbar}-i \theta}$  with an anisotropic scaling limit by deforming the connection $A$ and $B$ field as
\begin{equation}
\begin{aligned}
A_r & \rightarrow A_r, \\
A_\theta & \rightarrow \hbar A_\theta, \\
B & \rightarrow \frac{1}{\hbar} B.
\end{aligned}
\end{equation}
As one full sends $\hbar\rightarrow 0$, the Lagrangian action for TBF theory is obtained as
\begin{equation}
S^I_{TBF}=\int_{\mathbb{TP}^1} d r d \theta \operatorname{Tr}\left[T\left(-\partial_\theta A_r\right)+B\left(\partial_r A_\theta+\left[A_r, A_\theta\right]\right)\right].
\end{equation}
For the sake of this discussion, we will refer to the aforementioned limit as the type I tropicalization.  One check that there is another inequivalent anisotropic scaling limit that one can take. We will refer this as type II tropicalization and will be defined by the scaling limit
\begin{equation}
\begin{aligned}
A_r & \rightarrow \hbar A_r, \\
A_\theta & \rightarrow  A_\theta, \\
B & \rightarrow \frac{1}{\hbar} B.
\end{aligned}
\end{equation}
The corresponding action for type II TBF theory is
\begin{equation}
S^{II}_{TBF}=\int_{\mathbb{TP}^1} d r d \theta  \operatorname{Tr}\left[T\left(\partial_r A_\theta\right)+B\left(-\partial_\theta A_r+\left[A_r, A_\theta\right]\right)\right].
\end{equation}
Why would one decide to employ the type I scaling limit as opposed to different variations? The answer is given to us by the underlying foliation structure given to us by the Maslov dequantization limit. In the tropical limit, the leaves of the foliation run along the $\theta$-direction and consequently, we want to look for constraints on the radial part of the connection $A_r$ that allow them to be projectable functions and/or 1-forms. This is precisely the condition $\partial_\theta A_r =0$.  Employing the type II tropicalization will not lead to basic/projectable forms but instead lead to objects that are constant transverse to the leaves of the foliation but can still vary along the leaves. This latter class of objects is unnatural from the point of view of foliated geometry. As a result, there is a unique tropicalization that is singled out by the foliation; we unambiguously refer to the unique type I tropicalization of BF theory as the TBF theory.

Type II TBF theory can actually be salvaged into a relevant theory by considering the opposite limit of a tropical limit which is known as an \textit{arctic limit}\footnote{We want to acknowledge that the term \textit{arctic limit} was first coined by a relative of Petr Ho\v rava immediately after the development of tropological sigma models, however, tropical semiring appeared before in mathematical literature \cite{Perrin1992}.}. The limit can be parametrized in terms as
\begin{equation}
z=e^{\hbar r+i \theta}, \quad \bar{z}=e^{\hbar r-i \theta}.
\end{equation}
In the limit as $\hbar\rightarrow 0$, the leaves of the foliation run along the $r$-direction and the direction that is transverse to the foliation is along the $\theta$-direction.  It was discussed in \cite{Albrychiewicz:2023ngk} that limit is more physically relevant for developing nonequilibrium string perturbation theory. The relevance to anisotropic topological field theories comes from the fact that the type II scaling limit is now the natural limit to take since the condition $\partial_r A_\theta=0$ characterizes a projectable/basic function when the leaves of the foliation now run along the $r$-direction.  As a result, the relevant anisotropic field theory is given to us by the type II articalization i.e.,
\begin{equation}
\begin{aligned}
A_r & \rightarrow \hbar A_r, \\
A_\theta & \rightarrow A_\theta, \\
B & \rightarrow \frac{1}{\hbar} B, \\
 z &=e^{\hbar r+i \theta}, \\
  \bar{z}&=e^{\hbar r-i \theta}.
\end{aligned}
\end{equation}
We call the resulting theory, the arctic BF theory or the \textit{ABF} theory for short. The Lagrangian action is
\begin{equation}
S_{ABF}=\int_{\mathbb{TP}^1} d r d \theta  \operatorname{Tr}\left[A\left(\partial_r A_\theta\right)+B\left(-\partial_\theta A_r+\left[A_r, A_\theta\right]\right)\right].
\end{equation}
Where now the tropical projective space is understood to have the same differential structure as $\mathbb{CP}^1$ but the leaves of the foliation now run along the $r$-direction. The $\theta$-direction is still periodic. Similarly to TBF theory, we unambiguously refer the type II articalization of BF theory as the ABF theory.

There are additional scaling limits that reduce down BF theory to some other class of theories but they are not anisotropic, in the sense that they treat the components of the connection $A$ isotropically, so we do not consider them as anisotropic topological field theories despite the Maslov dequantization guaranteeing that they live on tropical projective space $\mathbb{TP}^1$. We are not aware of any sort of scenario where they might be relevant at the moment. Nonetheless, we will discuss them for the sake of completeness. The Type III scaling limit is
\begin{equation}
\begin{aligned}
A_r & \rightarrow \hbar A_r, \\
A_\theta & \rightarrow \hbar  A_\theta, \\
B & \rightarrow \frac{1}{\hbar} B.
\end{aligned}
\end{equation}
This leads to the action
\begin{equation}
S^{\text {III }}_{TBF}=\int d r  d \theta  \operatorname{Tr [}B\left(\partial_r A_\theta-\partial_\theta A_r\right)].
\end{equation}
This is an unusual theory due to the fact the Lie-algebra associated to the gauge group $G$ is still generically non-abelian but the structure constants do not explicitly enter into the action. At the level of the action, the lie-algebra components of the connection are independent. The type IV scaling limit is
\begin{equation}
\begin{aligned}
A_r & \rightarrow \hbar A_r, \\
A_\theta & \rightarrow \hbar  A_\theta, \\
B & \rightarrow \frac{1}{\hbar^2} B.
\end{aligned}
\end{equation}
Here, the action that we obtain is
\begin{equation}
S_{I V}=\int d r   d \theta \operatorname{Tr}[B\left(\left[A_r, A_\theta\right]\right)+T\left(\partial_r A_\theta-\partial_\theta A_r\right)].
\end{equation}
Integrating out the $B$ field imposes a sort of abelianization condition
\begin{equation}
B^a A_r^b A_\theta^c f_{b c}^a=0,
\end{equation}
here, the latin indices ${a,b,c}$ refer to Lie-algebra adjoint indices and $f^{a}_{bc}$ are the structure constants. We emphasize again that these last two scaling limits are not resulting in anisotropic field theories, so we don't really consider them as relevant for this paper.

\bibliographystyle{JHEP}
\bibliography{Bibliography/0CAnisotropic}

\end{document}